\documentclass[a4paper,11pt]{article}
\usepackage{amsmath,amssymb,color,comment}
\usepackage{slashed, tensor, bm, physics}
\usepackage{caption}
\usepackage{graphicx}
\usepackage{multirow}
\usepackage{enumerate}
\usepackage{subcaption}
\usepackage[normalem]{ulem}
\usepackage{float}
\usepackage[compat=1.1.0]{tikz-feynhand}
\usepackage{jheppub}

\usepackage[T1]{fontenc} 
\usepackage{booktabs} 
\title{
Excluding MeV-scale QCD axions by $K_L \to \pi^0\pi^0 a$ at KTeV
}
\author[a]{Takaya Iwai,}
\author[a]{Ryosuke Sato,}
\author[b,c,d]{Kohsaku Tobioka,}
\author[a]{and Takumu Yamanaka}

\affiliation[a\,]{Department of Physics, The University of Osaka, Toyonaka, Osaka 560-0043, Japan}
\affiliation[b\,]{Department of Physics, Florida State University, 77 Chieftan Way, Tallahassee, FL 32306, USA}
\affiliation[c\,]{Theory Center, High Energy Accelerator Research Organization (KEK), 1-1 Oho, Tsukuba, Ibaraki 305-0801, Japan}
\affiliation[d\,]{Yukawa Institute for Theoretical Physics, Kyoto University, Kyoto 606-8502, Japan}

\emailAdd{t\_iwai@het.phys.sci.osaka-u.ac.jp}
\emailAdd{rsato@het.phys.sci.osaka-u.ac.jp}
\emailAdd{ktobioka@fsu.edu}
\emailAdd{yamanaka@het.phys.sci.osaka-u.ac.jp}

\abstract{
An interesting proposal suggests that a QCD axion $(a)$ coupling to the up quark, down quark, and electron remains viable for an axion mass near 10~MeV. In this paper, this possibility is reexamined by deriving new bounds from kaon decays. In particular, we perform a detailed analysis of the $K_L \to \pi^0 \pi^0 e^+ e^-$ measurement reported by the KTeV experiment, and reinterpret $K^+$ decay measurements at the E949 and NA62 experiments to constrain both the diphoton decay and effectively invisible decay modes of the axion. 
We find that, combined with the previously known bounds, the viable window for the MeV-scale QCD axion is excluded, primarily due to the KTeV bound. Uncertainties associated with the chiral Lagrangian are further examined, and the scenario remains excluded even after accounting for these uncertainties, except for a tiny region of parameter space where higher-order corrections must finely cancel the leading-order contribution, suppressing the branching ratio of $K_L\to \pi^0\pi^0 a$ by three orders of magnitude.
}

\begin{document} 
\begin{flushright}
OU-HET-1301
\end{flushright}
\maketitle
\flushbottom

\section{Introduction}
The QCD axion \cite{Peccei:1977hh, Peccei:1977ur, Weinberg:1977ma, Wilczek:1977pj} is one of the leading candidates for the solution to the strong CP problem \cite{Jackiw:1976pf, Callan:1976je}.
It is a pseudo Nambu-Goldstone boson associated with the spontaneous breaking of $U(1)_{\rm PQ}$ Peccei-Quinn (PQ) symmetry, and the QCD $\theta$-term is absorbed at the QCD axion potential minimum \cite{Vafa:1984xg}. As a result, even though the CP-violating phase in Cabibbo-Kobayashi-Maskawa matrix is ${\cal O}(1)$, the neutron electric dipole moment is predicted to be small enough \cite{Georgi:1986kr} compared to the current experimental bound \cite{Abel:2020pzs}.
In the original PQWW axion model \cite{Peccei:1977hh, Peccei:1977ur, Weinberg:1977ma, Wilczek:1977pj}, the QCD axion decay constant $f_a$ is around the electroweak scale, and this model has been excluded by the measurement of the rare meson decay etc \cite{Kim:1986ax, Cheng:1987gp}. Thus, the QCD axion with higher $f_a$ \cite{Kim:1979if,Shifman:1979if,Dine:1981rt,Zhitnitsky:1980tq} has been extensively studied, and the current conventional wisdom tells us that $f_a$ should be larger than $\sim 10^8$ GeV. For a review, see, e.g., refs.~\cite{Kim:2008hd, Caputo:2024oqc}.

However, in 2017, Alves and Weiner \cite{Alves:2017avw} pointed out that an interesting viable scenario with $f_a$ as low as 1 GeV had been overlooked. Using the axion mass formula $m_a \sim f_\pi m_\pi/f_a$ where $f_\pi \simeq 92~{\rm MeV}$, this $f_a$ corresponds to an axion mass around 10 MeV. 
In this scenario, referred to as the MeV axion model in this paper, the axion couples only to the first generation of quarks and electrons.
Since the $a\to e^+e^-$ decay mode makes the axion lifetime as short as $\tau \lesssim 10^{-13}$~s, the stringent bound on $K^+ \to \pi^+ a$ with $a$ being invisible, as well as the bounds from beam-dump experiments, can be relaxed.
Although measurements of $K^+ \to \pi^+ e^+ e^-$ could constrain the QCD axion via $K^+ \to \pi^+ a$ with the axion visible decay $a\to e^+ e^-$, the specific choice of the PQ charges for the up and down quarks suppresses the mixing between the axion and the neutral pion, and it was claimed that the exclusion from the measurements of $K^+ \to \pi^+ a$ with $a \to e^+ e^-$ is not conclusive. The upper bound on the axion mass is derived using $B^0 \to K^{*0} e^+ e^-$ at the LHCb experiment \cite{Girmohanta:2024nyf, LHCb:2013pra, LHCb:2015ycz}, and the currently viable mass range is $2 m_e\leq m_a\leq 30$~MeV.
Other related discussions can be found in refs.~\cite{Alves:2020xhf, Liu:2021wap}.

In this paper, we reexamine the experimental constraints on the MeV axion model~\cite{Alves:2017avw}. In particular, we scrutinize the constraint on $K_L \to \pi^0 \pi^0 a$ with $a \to e^+ e^-$ derived from the existing measurement of $K_L \to \pi^0 \pi^0 e^+ e^-$. The upper bound on ${\rm Br}(K_L \to \pi^0 \pi^0 e^+ e^-)$ was reported as $6.6 \times 10^{-9}$ by the KTeV experiment~\cite{KTeV:2002tpo}. Although this KTeV result was briefly discussed in the context of the MeV axion model~\cite{Goudzovski:2022vbt}, using the rate of ${\rm Br}(K_L \to \pi^0 \pi^0 a)$ calculated in ref.~\cite{Alves:2020xhf}, a dedicated analysis of the corresponding constraint has not been performed.

While the KTeV collaboration analyzed the signal based on the Sehgal model~\cite{Heiliger:1993qi} to evaluate the Standard Model contribution, we reinterpret the result in the context of the MeV axion by performing a Monte Carlo simulation of the signal process that accounts for the detector geometry and resolution. We also derive the constraints from $K^+$ decay measurements at the NA62 and E949 experiments using a conservative estimate of the $K^+ \to \pi^+ a$ decay rate, and consider other relevant limits, such as the one from the electron anomalous magnetic moment measurement.  As we will show, the MeV axion scenario is generically excluded by these measurements, unless higher-order effects lead to a significant cancellation of the relevant production rates.

The outline of this paper is as follows. In section \ref{sec:MeVaxion}, we briefly review the MeV axion, discuss the production rate and bounds for $K^+\to \pi^+ a$, and calculate ${\rm Br}(K_L \to \pi^0 \pi^0 a)$. In section \ref{sec:KTeV}, we discuss the constraint from the KTeV experiment. In section \ref{sec:result}, we show the constraints from the KTeV experiment on the parameter space of the MeV axion model together with the existing constraints. Finally, we conclude in section \ref{sec:Conclusion}.

\section{MeV axion model and \texorpdfstring{$K_L \to \pi^0 \pi^0 a$}{kl to pi0 pi0 a}}\label{sec:MeVaxion}
\begin{figure}[tbp]
    \centering
    \includegraphics[width=0.86\linewidth]{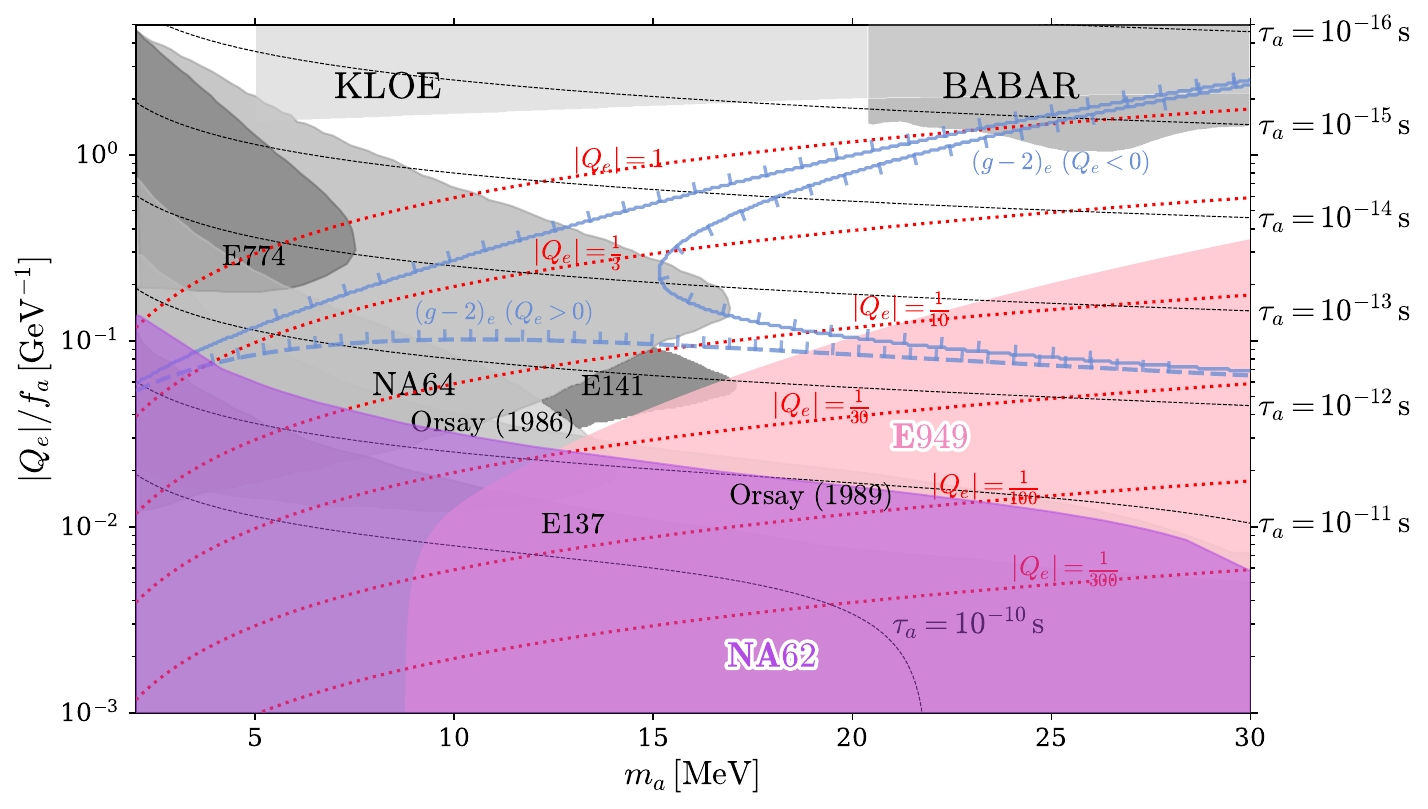}
    \includegraphics[width=0.86\linewidth]{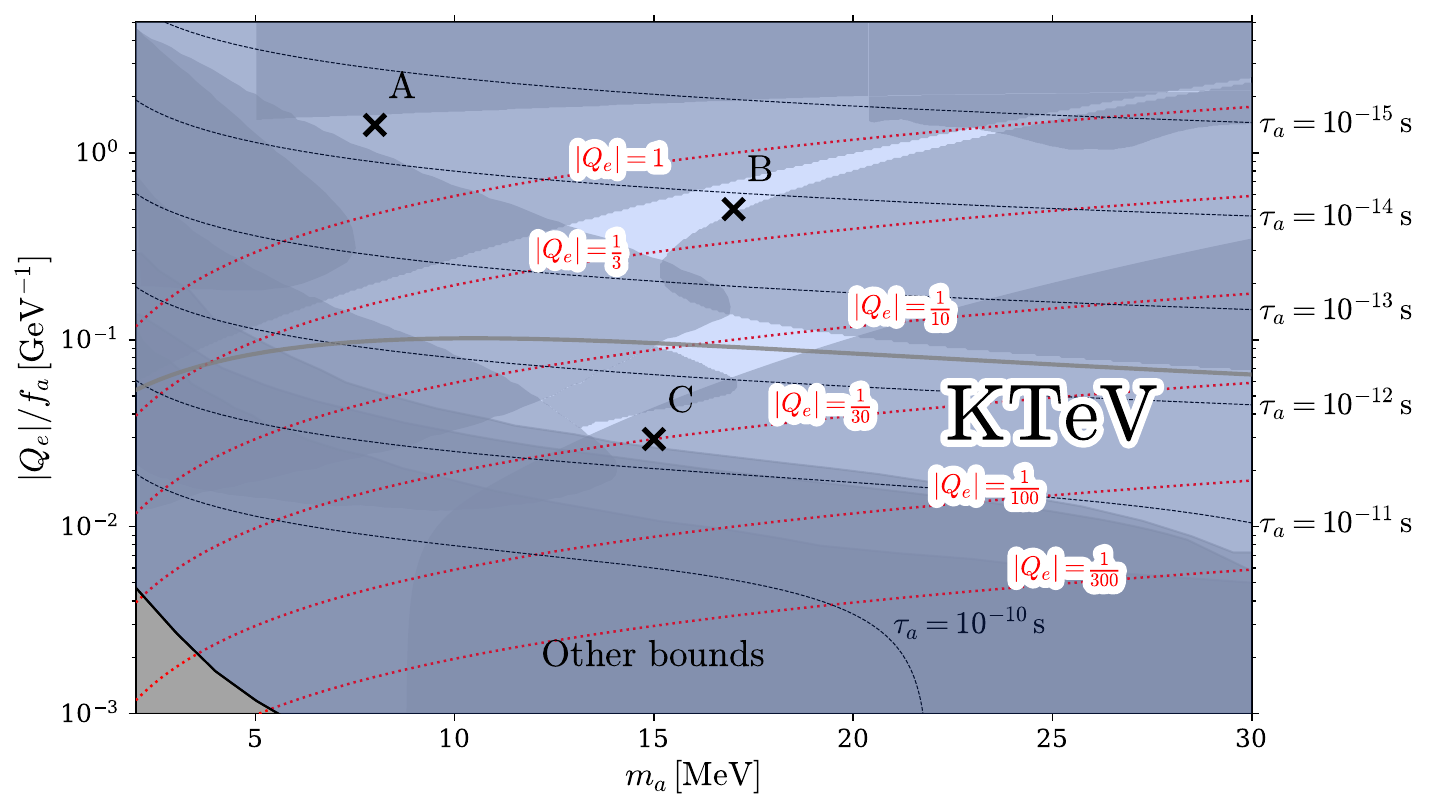}
    \caption{\textit{Top}:
    Summary of current experimental constraints on our axion scenario in the $(m_a,\,|Q_e|/f_a)$ plane.
    The red curves show contours of constant $|Q_e|$,
    while the black dashed curves represent constant axion lifetimes. The shaded gray regions correspond to the previously existing exclusion limits from electron beam-dump experiments and
    $e^+ e^-$ collider experiments \cite{Bross:1989mp,Riordan:1987aw,Davier:1986qq,Andreev:2021fzd,NA64:2021aiq,Gninenko:2017yus,CCM:2021jmk,Bjorken:1988as,Davier:1989wz,BaBar:2014zli,Anastasi:2015qla}.
    In contrast, the purple region indicates the NA62 constraint 
    \cite{NA62:2025upx}, while the shaded pink region shows the bound derived in this work from the experimental limit on the radiative kaon decay ${\rm Br}(K^+\to\pi^+\gamma\gamma)\lesssim 5.0\times10^{-7}$ set by E949. The blue solid (dashed) curve shows the constraint from the electron anomalous magnetic moment, obtained by requiring $|\Delta a_e|<10^{-12}$ for $Q_e>0$ ($Q_e<0$).
    \textit{Bottom}: Same as in the top panel, with the additional constraint from the KTeV experiment derived in our work. The KTeV bound is derived employing the axion mixing with $\eta_{ud}$ and $\eta_{s}$ at the leading order in the chiral Lagrangian. The hatched region above the black solid line is excluded by the KTeV result. The gray contour indicates the upper limit from the electron anomalous magnetic moment for $Q_e>0$, obtained by requiring $|\Delta a_e|<10^{-12}$. Three benchmark points $(m_a, Q_e)$ adopted in our analysis are highlighted by black crosses.
    }

    \label{fig:summary-plot}
\end{figure}

In this section, we briefly review the MeV axion model~\cite{Alves:2017avw} and calculate the branching ratio of $K_L \to \pi^0 \pi^0 a$.
Although it has been commonly believed that $\mathcal{O}(10)\,{\rm MeV}$ axion models are already excluded by the measurement of the rare meson decay etc \cite{Kim:2008hd}, ref.~\cite{Alves:2017avw} found that it is possible to evade the existing constraints by imposing certain conditions on the model. 
These conditions are as follows:
\begin{enumerate}[(i)]
\item the axion couples only to the first-generation fermions, including electrons.
\item the ratio of the PQ charges of the up and down quarks is taken to be $2:1$.
\item the axion to electron charge $Q_e$ is introduced to avoid stringent bounds from $K^+\to \pi^+ X_{\rm inv}$ where $X_{\rm inv}$ represents an invisible particle.
\end{enumerate}
From the condition (i), the axion does not couple to heavier generations further, and the MeV axion model evades the limits from quarkonium decays ($J/\psi,\,\Upsilon \to a\gamma$) and from the muon anomalous magnetic moment  $(g-2)_\mu$.
Also, from the condition (ii), the mixing between the axion and the neutral pion is suppressed and the constraints from the measurement on $\pi^+ \to e^+ \nu_e a$ and $K^+ \to \pi^+ a$ are relaxed as we will discuss in the following.

According to the recipe described above, we obtain the low-energy effective Lagrangian of the model below the PQ breaking scale $f_a$ as
\begin{align}
    \mathcal{L}^{\rm eff}_a=\frac{m_u}{f_a}iQ_ua\bar{u}\gamma_5u+\frac{m_d}{f_a}iQ_da\bar{d}\gamma_5d+\frac{m_e}{f_a}iQ_ea\bar{e}\gamma_5e.
    \label{eq:Leff_axion}
\end{align}
In the following, we take the PQ charge of the up and down quark charges as $Q_u = 2$ and $Q_d = 1$. As we will see later, this ratio $Q_u/Q_d = 2$ suppresses the mixing between the axion and the neutral pions, and the MeV axion model evades severe experimental constraints.
Then, we have only two remaining parameters: the PQ breaking scale $f_a$ and the electron PQ charge $Q_e$.
Examples of UV completion above the PQ breaking $f_a$ and the electroweak scale are given in refs.~\cite{Alves:2017avw, Liu:2021wap, Girmohanta:2024nyf}.

Figure \ref{fig:summary-plot} shows a summary of the status of the MeV axion model based on the leading order chiral Lagrangian.
The top panel shows a summary of the current experimental constraints on the MeV axion model.
Details of the chiral Lagrangian and the experimental constraints will be explained in this section.
The bottom panel shows a constraint from the KTeV experiment, which will be discussed in section \ref{sec:KTeV} and \ref{sec:result}.

\subsection{Effective Lagrangian for the MeV axion} \label{sec:axion EFT}
The quark-level axion couplings in eq.~(\ref{eq:Leff_axion}) is related to the hadronic observables by the chiral Lagrangian as
\begin{align}
    {\cal L} = \frac{f_\pi^2}{4} {\rm tr}[(\partial_\mu U)(\partial^\mu U^\dagger)] + \frac{1}{2}f_\pi^2 B_0 {\rm tr}[M_q U^\dagger + M_q^\dagger U] - \frac{1}{2}M_0^2 \eta_0^2, \label{eq:chpt}
\end{align}
where $f_\pi \simeq 92~\mathrm{MeV}$ is the pion decay constant and $M_q$ is given by
\begin{align}
M_q = {\rm diag}\left( m_u \exp\left( \frac{-iQ_u a}{f_a} \right),~m_d \exp\left( \frac{-iQ_d a}{f_a}\right),~m_s \right).
\end{align}
The meson field $U$ is defined as
\begin{align}
    U = \exp\left( \frac{\sqrt{2}i\Phi}{f_\pi} \right), \quad
\Phi =
\begin{pmatrix}
\dfrac{\pi^0}{\sqrt{2}} + \dfrac{\eta_8}{\sqrt{6}} + \dfrac{\eta_0}{\sqrt{3}} & \pi^+ & K^+ \\[6pt]
\pi^- & -\dfrac{\pi^0}{\sqrt{2}} + \dfrac{\eta_8}{\sqrt{6}} + \dfrac{\eta_0}{\sqrt{3}} & K^0 \\[6pt]
K^- & \bar{K}^0 & -\dfrac{2\,\eta_8}{\sqrt{6}} + \dfrac{\eta_0}{\sqrt{3}}
\end{pmatrix}.
\label{eq:Phi_matrix_U3}
\end{align}

By using the effective Lagrangian in eq.~\eqref{eq:chpt}, we can calculate the axion mass $m_a$ as
\begin{align}
    m_a
    &= \frac{3}{\sqrt{1+\epsilon_{\eta\eta'}}}\,
      \frac{\sqrt{m_u m_d}}{m_u+m_d}\,
      \frac{m_\pi f_\pi}{f_a} 
    \simeq 17~{\rm MeV} \times \left( \frac{f_a}{1~{\rm GeV}} \right)^{-1}, \label{eq:axion mass}
\end{align}
where $\epsilon_{\eta\eta'}$ represents a correction from $\eta$ and $\eta'$ as
\begin{align}
    \epsilon_{\eta\eta'}
    = \frac{m_u m_d}{m_s(m_u+m_d)}\left( 1 + 6 \frac{B_0 m_s}{M_0^2} \right) \simeq 0.04.
\end{align}
Here we have used $B_0 m_s \simeq m_K^2$, $M_0^2\simeq m_{\eta'}^2$.
The mixing angles between the axion and neutral mesons, $\theta_{a\pi}$, $\theta_{a\eta_8}$, and $\theta_{a\eta_0}$ can be obtained from the leading order chiral Lagrangian given in eq.~\eqref{eq:chpt} as\footnote{
In general, the mixing angle between the axion and neutral meson has an ambiguity under a chiral phase redefinition of the quark fields as pointed out in ref.~\cite{Bauer:2021wjo}.
In the following, we fix a definition of the axion and the quark field so that the axion does not have derivative coupling with quarks to avoid this ambiguity.}
\begin{align}
    \theta_{a\pi}^{\rm (LO)} &= -\frac{f_\pi}{f_a} \frac{1}{1 + \epsilon_{\eta\eta'}} \left( \frac{Q_u m_u - Q_d m_d}{m_u+m_d} + \epsilon_{\eta\eta'} \frac{Q_u- Q_d}{2} \right) , \\
    \theta_{a\eta_8}^{\rm (LO)} &= -\frac{\sqrt{3}}{2} \frac{f_\pi}{f_a} \frac{\epsilon_{\eta\eta'}}{1 + \epsilon_{\eta\eta'}} \frac{1 + 2B_0m_s / M_0^2}{1 + 6B_0m_s / M_0^2} (Q_u+Q_d), \\
    \theta_{a\eta_0}^{\rm (LO)} &= -\frac{\sqrt{6}}{Q_u+Q_d} \frac{f_a}{f_\pi} \frac{m_a^2}{M_0^2}.
\end{align}
Numerical value of the mixing angles are obtained as
\begin{align}
    \theta_{a\pi}^{\rm (LO)} &\simeq (2.8 \pm 1.6) \times 10^{-3} \times \left( \frac{f_a}{1~{\rm GeV}}\right)^{-1}, \label{eq:a pi mixing}\\
    \theta_{a\eta_8}^{\rm (LO)} &\simeq -5.4 \times 10^{-3} \times \left( \frac{f_a}{1~{\rm GeV}}\right)^{-1},\label{eq:a eta8 mixing}\\
    \theta_{a\eta_0}^{\rm (LO)} &\simeq -2.8 \times 10^{-3} \times \left( \frac{f_a}{1~{\rm GeV}}\right)^{-1}. \label{eq:a eta0 mixing}
\end{align}
In the above evaluation, we only show the error from the PDG 2025 light quark mass ratio $m_u/m_d = 0.462 \pm 0.013$   \cite{ParticleDataGroup:2024cfk}\footnote{Ref.~\cite{Alves:2017avw} used PDG 2017 the light quark mass ratio $m_u/m_d = 0.483 \pm 0.027$
\cite{ParticleDataGroup:2016lqr}, which is a weighted average value of refs.~\cite{Blum:2007cy,MILC:2009mpl,Blum:2010ym,Aoki:2012st,EuropeanTwistedMass:2014osg,MILC:2015ypt,Fodor:2016bgu}, and obtained as
\begin{align}
    \theta_{a\pi}^{\rm (LO)} = (0.2 \pm 3) \times 10^{-3} \times \left( \frac{f_a}{1~{\rm GeV}}\right)^{-1}. \nonumber
\end{align}
Although PDG 2025 $m_u/m_d$ does not include a new data which has not been included in PDG 2017, refs.~\cite{Blum:2007cy, Aoki:2012st} is not used in PDG 2025.} and other errors are neglected.
Because of a cancellation in $Q_u m_u - Q_d m_d$ in $\theta_{a\pi}^{\rm (LO)}$, we observe $\theta_{a\pi}^{\rm (LO)}$ becomes the same order of $\theta_{a\eta_8}^{\rm (LO)}$ and $\theta_{a\eta_0}^{\rm (LO)}$ or can be even smaller than them.
For later discussions, we introduce $\eta_{ud}$ and $\eta_s$ fields defined as
\begin{align}
    \left(\begin{array}{c}
    \eta_8 \\ \eta_0
    \end{array}\right)
    =
    \left(\begin{array}{cc}
    \sqrt{1/3} & -\sqrt{2/3} \\
    \sqrt{2/3} & \sqrt{1/3}
    \end{array}\right)
    \left(\begin{array}{c}
    \eta_{ud} \\ \eta_s
    \end{array}\right).
\end{align}
The axion mixing angles $\theta_{a\eta_{ud}}^{\rm (LO)}$ and $\theta_{a\eta_s}^{\rm (LO)}$ are given as
\begin{align}
    \theta_{a\eta_{ud}}^{\rm (LO)} &= \sqrt{\frac{2}{3}} \theta_{a\eta_8}^{\rm (LO)} + \sqrt{\frac{1}{3}} \theta_{a\eta_0}^{\rm (LO)} \simeq -6.0 \times 10^{-3} \times \left( \frac{f_a}{1~{\rm GeV}}\right)^{-1},\label{eq:a etaud mixing LO}
    \\
    \theta_{a\eta_{s}}^{\rm (LO)} &= -\sqrt{\frac{1}{3}} \theta_{a\eta_8}^{\rm (LO)} + \sqrt{\frac{2}{3}} \theta_{a\eta_0}^{\rm (LO)} \simeq 8.5 \times 10^{-4} \times \left( \frac{f_a}{1~{\rm GeV}}\right)^{-1}. \label{eq:a etas mixing LO}\end{align}

By integrating out mesons, we obtain the effective interaction with the axion and photon as
\begin{align}
    {\cal L} = g_{a}^{\gamma\gamma} a F_{\mu\nu} \tilde F^{\mu\nu},
\end{align}
where
\begin{align}
g^{\gamma\gamma}_{a}
   = \frac{\alpha}{4\pi f_{\pi}}
     \left(
       \theta_{a\pi} + 
       \frac{5}{3}\,\theta_{a\eta_{ud}}
       + \frac{\sqrt{2}}{3}\,\theta_{a\eta_{s}}
     \right).
\label{gagg}
\end{align}
For $m_a > 2m_e$, the axion can decay into $e^+ e^-$ and $\gamma\gamma$ and its decay widths are
\begin{align}
    \Gamma(a \to e^+ e^-) &= \frac{m_a}{8\pi} \left( \frac{Q_e m_e}{f_a} \right)^2 \sqrt{ 1-\frac{4m_e^2}{m_a^2} }, \\
    \Gamma(a \to \gamma\gamma) &= \left( g_a^{\gamma\gamma}\right)^2 \frac{m_a^3}{4\pi}.
    \label{axiondecay}
\end{align}
Note that electron loop diagram contribution is negligible in $\Gamma(a\to\gamma\gamma)$ in a parameter region with our interest \cite{Alves:2017avw}.
The axion lifetime is given by
\begin{equation}
\tau_a
= \frac{1}{\Gamma(a \to e^+ e^-) + \Gamma(a \to \gamma\gamma)}.
\label{zyumyou}
\end{equation}
For $Q_e \gtrsim 1.7\times10^{-3}\left( m_a/10~{\rm MeV} \right)$, 
the axion dominantly decays into $e^+ e^-$, and, in this regime, the lifetime is approximated by
\begin{align}
    \tau_a
    &= \left( \frac{m_a}{8\pi}\right)^{-1} \left( \frac{Q_e m_e}{f_a} \right)^{-2} \left( 1-\frac{4m_e^2}{m_a^2}\right)^{-1/2} \nonumber\\
    &\simeq 1.8\times 10^{-14}~{\rm s}\times \frac{1}{Q_e^2} \left( \frac{m_a}{10~{\rm MeV}} \right)^{-3} \left( 1-\frac{4m_e^2}{m_a^2}\right)^{-1/2}. \label{eq:axion lifetime}
\end{align}
In addition, since $m_a > 30~{\rm MeV}$ has been excluded by the measurement of $B^0 \to K^{*0} e^+ e^-$ at the LHCb experiment \cite{Girmohanta:2024nyf, LHCb:2013pra, LHCb:2015ycz}, we focus on the mass range $2m_e < m_a < 30~{\rm MeV}$ in what follows.

In the MeV axion model, the axion-pion mixing $\theta_{a\pi}$ is severely constrained from a measurement of $\pi^+ \to  e^+ \nu_e e^+ e^-$ at the SINDRUM experiment.
It induces the decay channel $\pi^+ \to e^+ \nu_e a$ with $a\to e^+ e^-$, whose decay width is given as \cite{Krauss:1986bq, Alves:2017avw}
\begin{align}
    \Gamma(\pi^+ \to e^+ \nu_e a) = \frac{\cos^2\theta_c G_F^2 m_\pi^2 \theta_{a\pi}^2}{384\pi^3},
\end{align}
where $\theta_c$ is the Cabibbo angle. 
The SINDRUM experiment put an upper bound on the branching fraction of this mode as $\sim 10^{-10}$ in the mass range $m_a \sim {\cal O}(1-10)$ MeV \cite{SINDRUM:1986klz}, and this value can be translated into the language of $\theta_{a\pi}$ as
\begin{align}
    |\theta_{a\pi}| \lesssim 10^{-4}. \label{eq:theta-pi}
\end{align}
Comparing this constraint with the leading order mixing angle $\theta_{a\pi}^{\rm (LO)}$ in eq.~\eqref{eq:a pi mixing}, we can find that this range is within $2\sigma$ range of $m_u/m_d$ value. In addition, this discussion suffers from ${\cal O}(1)$ uncertainty from the next-leading order contribution in the chiral Lagrangian. Thus, the constraint from the SINDRUM experiment is not conclusive for the MeV axion model.
In the following analysis, we assume that the axion is pion-phobic and the mixing angle is suppressed as $|\theta_{a\pi}| < 10^{-4}$ to be consistent with the constraint from the SINDRUM experiment. 

Since the viable MeV axion is pion-phobic, the axion mainly mixes with $\eta$ and $\eta'$. However, the mixing angles $\theta_{a\eta_8}$ and $\theta_{a\eta_0}$ suffer from ${\cal O}(1)$ uncertainty because the next-leading order correction in the chiral Lagrangian is of the order of $(m_K/4\pi f_\pi)^2,$ which is less than but close to 1.
Ref.~\cite{Alves:2017avw} utilizes the chiral Lagrangian in ref.~\cite{Leutwyler:1997yr} which is originally developed to discuss $\eta$-$\eta'$ mixing, and obtained 
as $\theta_{a\eta_{ud}} \simeq -1.9\times 10^{-3} \times \left(f_a/1~{\rm GeV}\right)^{-1}$.
This value is $\sim 1/3$ of $\theta_{a\eta_{ud}}^{\rm (LO)}$ and we can see $\theta_{a \eta_{ud}}$ obtains ${\cal O}(1)$ correction for the leading-order value.
Note that there exist next-leading order terms in the chiral Lagrangian which are not taken into account in the above evaluation, and the most general chiral Lagrangian has too many parameters to be determined from the low energy observables.
In the following paper, we do not determine the mixing angle at the next-leading order but utilize the leading order mixing angle  $\theta_{a\eta_{ud}}^{\rm (LO)}$ and $\theta_{a\eta_s}^{\rm (LO)}$ while indicating ${\cal O}(1)$ uncertainty.

\subsection{Decay of kaons and \texorpdfstring{$\Delta S = 1$}{delta s = 1} interactions}
Measurements of the rare decays of kaons are sensitive to the MeV axion.
To discuss decay of $K^+$ and $K_L$ involving the axion, we introduce $\Delta S = 1$ interaction terms in the effective Lagrangian, and calculate $K^+ \to \pi^+ a$ and $K_L \to \pi^0 \pi^0 a$.

\subsubsection{Octet enhancement and \texorpdfstring{$K^+ \to \pi^+ a$}{k+ to pi+ a}}
First, we discuss $K^+ \to \pi^+ a$. 
This decay channel is described by the effective interaction for $\Delta S = 1$, and the effective interaction can be classified into $SU(3)$ flavor octet and 27-plet.
The leading order term is
\begin{align}
    \mathcal{L}_8^{(\Delta S=1)} 
    &= g_8 f_\pi^2 \, {\rm Tr}\!\left( \lambda_{ds} \, \partial_\mu U \, \partial^\mu U^\dagger \right) + \text{h.c.}, \label{eq:O8} \\[4pt]
    \mathcal{L}_{27}^{(\Delta S=1)} 
    &
= g_{27} f_\pi^2
t^{ij}_{kl} (U^\dagger \partial_\mu U)^k_{~i} (U^\dagger \partial^\mu U)^l_{~j}.
\label{eq:O27-trace}
\end{align}
We define 
$\lambda_{ds} \equiv (\lambda_6 + i\lambda_7)/2$, 
and the coefficients $t^{ij}_{kl}$ denote the Clebsch-Gordan factors that project onto the 27-plet representation \cite{Kambor:1989tz, Alves:2020xhf}.
We also discuss a next-leading order octet operators discussed in ref.~\cite{Alves:2017avw}:
\begin{align}
    \mathcal{L}_8'^{(\Delta S=1)} 
    &= -\,\frac{2g_8' f_\pi^2 B_0}{\Lambda_\chi^2}
    \,{\rm Tr}\!\left(\lambda_{ds} M_q^\dagger U^\dagger \right)
    {\rm Tr}\!\left(\partial_\mu U \, \partial^\mu U^\dagger\right)
    + \text{h.c.}. \label{eq:O8prime}
\end{align}
Here, $\Lambda_\chi \sim 4\pi f_\pi$ is the chiral Lagrangian cutoff scale.

It is well known that there exists a large hierarchy between the decay width of $K_S \to \pi\pi$ and $K^+ \to \pi^+ \pi^0$ as \cite{ParticleDataGroup:2024cfk}
\begin{align}
    \frac{ \Gamma(K_S \to \pi\pi) }{ \Gamma(K^+ \to \pi^+ \pi^0) } \simeq 668. \label{eq:ratio ks k+}
\end{align}
Although $K_S \to \pi\pi$ is contributed from both 27-plet and octet operators, $K^+ \to \pi^+ \pi^0$ is contributed only from 27-plet operator. This indicates a hierarchy between the coefficient of the octet and 27-plet operators \cite{Donoghue:1992dd}. Actually, the ratio in eq.~\eqref{eq:ratio ks k+} is translated into language of $g_8$, $g'_8$, and $g_{27}$ as
\cite{Alves:2017avw, Alves:2020xhf}
\begin{align}
    g_{27} \simeq 0.032 \times \left| g_8 + g'_8\left(  \frac{2m_K}{\Lambda_\chi} \right)^2 \right|.
\end{align}
Thus, $K \to \pi\pi$ tells us that a linear combination of $g_8$ and $g'_8$ are much larger than $g_{27}$, however, it cannot disentangle $g_8$ and $g'_8$.

On the other hand, a hierarchy between $g_8$ and $g'_8$ can significantly change the decay width of $K^+ \to \pi^+ a$.
Both eq.~\eqref{eq:O8} and eq.~\eqref{eq:O8prime} do not involve the axion field directly, the contribution to $K$ decay with the axion comes from the axion mixing with $\eta_{ud}$.
The operator $g_8$ contributes to $K^+ \to \pi^+ a$ via $a$-$\eta_{ud}$ mixing although $g'_8$ does not.
In the case of $g_8 \gg g_{27} \sim g'_8$, we obtain \cite{Alves:2017avw}
\begin{align}
    {\rm Br}(K^+ \to \pi^+ a) |_{g_8 \gg g_{27}} \simeq 4 \times 10^{-3} \times \left| \frac{\theta_{a\eta_{ud}}}{ 5\times 10^{-3} } \right|^2. \label{eq:Kpia g8}
\end{align}
Comparing this number with eq.~\eqref{eq:a etaud mixing LO},  we can see that $f_a = {\cal O}(1)~{\rm GeV}$ has been already severely constrained.
On the other hand, in the case of $g'_8 \gg g_8 \sim g_{27}$, the branching ratio becomes small.
If the contribution from $g_{27}$ operator dominates, we obtain \cite{Alves:2020xhf}
\begin{align}   
{\rm Br}(K^+\to\pi^+a)|_{g_{27}} \simeq
2\times10^{-6} \times \left|
  \frac{\theta_{a\eta_{ud}} + (\sqrt{2}/3) \theta_{a\eta_{s}}}
       {5\times10^{-3}}
\right|^{2}. \label{eq:Kpia g27}
\end{align}

%%%%%%%%%%%%%%%%%%%%%%%%%%%%%%%%%%
\subsubsection{Bounds on \texorpdfstring{$K^+ \to \pi^+ a$}{k+ to pi+ a}}
\label{Kplus}

The two possibilities for octet enhancement discussed above can be distinguished by considering the measurement of \(K^+ \to \pi^+ e^+ e^-\), since axions prefer to decay into \(e^+ e^-\) with relatively short lifetimes. In ref.~\cite{Alves:2017avw}, the BNL measurement of \(K^+ \to \pi^+ e^+ e^-\)~\cite{Baker:1987gp} was reexamined, leading to an upper bound on \({\rm Br}(K^+ \to \pi^+ a)\) at the level of \(10^{-5}\).  This bound constrains the mixing angles \(\theta_{a\eta_{ud}}\) and \(\theta_{a\eta_s}\), assuming that \(\theta_{a\pi}\) is negligible to evade the SINDRUM bound, as in eq.~\eqref{eq:theta-pi}.

In the case \(g_8 \gg g_{27} \sim g'_8\), \({\rm Br}(K^+ \to \pi^+ a)\) is enhanced by the octet operator, as in eq.~\eqref{eq:Kpia g8}, resulting in stringent constraints. On the other hand, if \(g'_8 \gg g_8 \sim g_{27}\), the octet enhancement is absent, and the branching ratio predicted by eq.~\eqref{eq:Kpia g27} might not have been excluded by the BNL measurement, as interpreted in ref.~\cite{Alves:2017avw}. Therefore, in the remainder of this paper, we focus on the latter viable scenario, \(g'_8 \gg g_8 \sim g_{27}\).

Even with this conservative estimate of the \(K^+ \to \pi^+ a\) production rate, we can still obtain several constraints. As the value of \(|Q_e|\) gets smaller, the branching ratio of \(a \to \gamma\gamma\) can be sizable and, if the axion lifetime is sufficiently short, measurements of \(K^+ \to \pi^+ \gamma \gamma\) exclude part of the parameter space. At the E949 experiment~\cite{E949:2005qiy}, an upper bound on the SM radiative kaon decay was obtained,
\begin{align}
{\rm Br}(K^+ \to \pi^+ \gamma \gamma)^{90\%\rm CL}_{\rm E949} = 8.5 \times 10^{-9}. \label{eq:E949bound_3body}
\end{align}
For axion searches, the kaon decay is a two-body process, \(K^+ \to \pi^+ a\), for which the acceptance is better than that of the three-body radiative decay. Therefore, the situation is expected to be closer to the \(K^+ \to \pi^+ \gamma\) search, and the corresponding upper bound at E949 is ${\rm Br}(K^+ \to \pi^+ \gamma)_{\rm E949}^{90\%\rm CL} = 2.3 \times 10^{-9}.$
Considering axion decays occurring inside the detector~\cite{Gori:2020xvq}, we derive the bound as
\begin{align}
     {\rm Br}(K^+ \to \pi^+ \gamma(\gamma))|_{\rm E949}^{90\%\rm CL}
     >
     {\rm Br}(K^+ \to \pi^+ a)|_{g_{27}}
     {\rm Br}(a\rightarrow\gamma\gamma)
     \left(1 - \exp\!\left[- L/(c\tau_a E_a/m_a)\right]\right),
     \label{eq:E949bound}
\end{align}
where we take a detector volume length \(L = 0.8\,\mathrm{m}\) and the typical axion energy of \(E_a = m_{K^+}/2\). Evaluating this constraint using the leading-order mixing angles in eqs.~\eqref{eq:a eta8 mixing} and \eqref{eq:a eta0 mixing}, the analysis using even the weaker upper bound of eq.~\eqref{eq:E949bound_3body} excludes a unique region of parameter space, as shown in pink in the top panel of figure~\ref{fig:summary-plot}. 

Furthermore, we examine the \(K^+ \to \pi^+ X_{\rm inv}\) bound using the NA62 measurement~\cite{NA62:2025upx}. This bound can be recast in the MeV axion model by estimating the fraction of axions that decay outside the detector,
\begin{align}
     {\rm Br}(K^+\to\pi^+X_{\rm inv})|_{\rm NA62}^{90\%\rm CL}
     >
     {\rm Br}(K^+\to\pi^+a)|_{g_{27}}
     \exp\!\left[-\lambda L/(c\tau_a E_a/m_a)\right],
     \label{eq:NA62bound}
\end{align}
where \({\rm Br}(K^+\to\pi^+X_{\rm inv})|^{90\%\rm CL}_{\rm NA62}\) is typically \(10^{-11}\) from figure~2(b) of ref.~\cite{NA62:2025upx}. We assume a typical axion energy of \(E_a = 30\,\mathrm{GeV}\) and a distance \(L = 150\,\mathrm{m}\) for the axion to escape the detector. Although both parameters vary in each \(K^+\) decay, we can reinterpret the bound by tuning a phenomenological parameter \(\lambda = 0.4\), such that the above formula reproduces the NA62 limits for lifetimes of $1\times 10^{-10}$\,s  and $2\times 10^{-10}$\,s~\cite{NA62:2025upx}. This NA62 search excludes  parameter space with axion lifetimes \(\tau_a \gtrsim 10^{-11}\,\mathrm{s}\), as shown in red in the top panel of figure~\ref{fig:summary-plot}.

\subsubsection{\texorpdfstring{$K_L \to \pi^0 \pi^0 a$}{kl to pi0 pi0 a}}
Next we discuss the decay mode $K_L \to \pi^0 \pi^0 a$. 
The KTeV experiment has searched for the decay $K_L \to \pi^0 \pi^0 e^+ e^-$ and has set an upper limit on the branching ratio $\mathrm{Br}(K_L \to \pi^0 \pi^0 e^+ e^-) < 6.6 \times 10^{-9}$ (90\% C.L.)~\cite{KTeV:2002tpo} for the Standard Model contribution based on the Sehgal model discussed in the appendix~\ref{sec:sehgal model}. 
Since the axion lifetime in eq.~\eqref{eq:axion lifetime} indicates that the axion decays promptly into an $e^+e^-$ pair, the KTeV measurement has sensitivity on the MeV axion model.

\begin{figure}[t]
    \centering
    \includegraphics[width=0.8\linewidth]{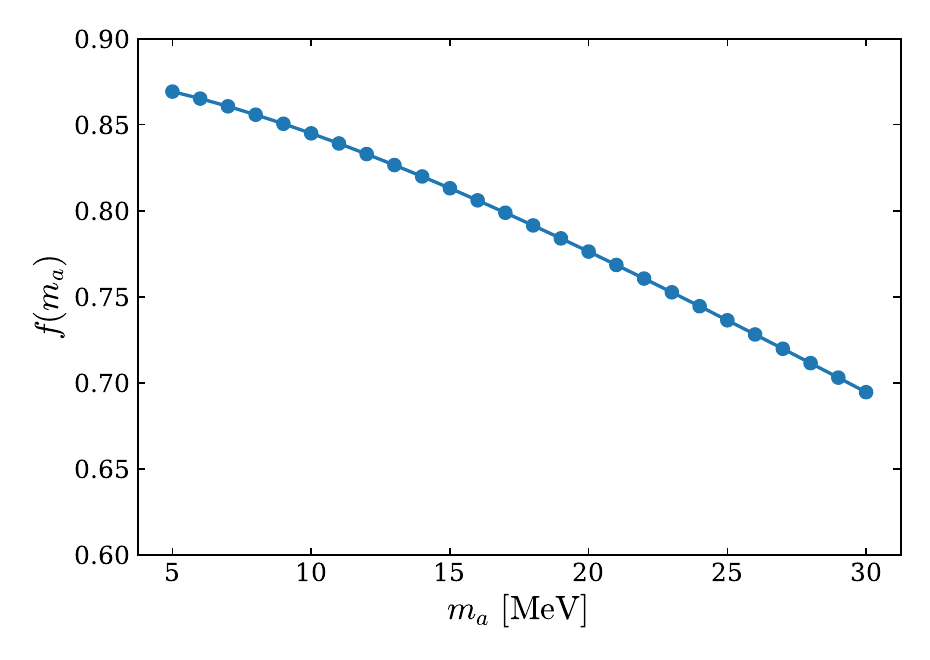}
    \caption{The function $f(m_a)$ encoding the $m_a$ dependence in ${\rm Br}(K_L \to \pi^0 \pi^0 a)$ given in eq.~\eqref{eq:BR_KLpi0pi0a}.}
    \label{BrKL}
\end{figure}

Let us evaluate ${\rm Br}(K_L\to \pi^0 \pi^0 a)$ by using the chiral Lagrangian described in the previous subsection. As we have discussed, we focus on the contribution from an effective interaction with $g'_8$ given in eq.~\eqref{eq:O8prime}. The detailed calculation involving nonzero $g_8$ and $g_{27}$ can be found in ref.~\cite{Alves:2020xhf}.
Since eq.~\eqref{eq:O8prime} does not include $a$ directly, the decay $K_L \to \pi^0 \pi^0 a$ is induced by the mixing of the axion with neutral pseudoscalar mesons $\varphi=\pi^0,~\eta_{ud},~\eta_s$.
We obtain the following amplitudes with off-shell $\pi^0$, $\eta_{ud}$, and $\eta_s$:
\begin{align}
    \mathcal{M}(K_L \!\to\! \pi^0 \pi^0 \pi^{0*})
    &= \frac{g'_8}{f_\pi^2} \left(\frac{2m_K}{\Lambda_\chi}\right)^2 \left( \frac{1}{3}m_K^2
 - \frac{4}{3}m_\pi^2 + \frac{1}{3}p_{\pi^0}^2 + m_\pi^2 Y \right), \label{eq:amp KL to pi pi pi}\\
    \mathcal{M}(K_L \!\to\! \pi^0 \pi^0 \eta_{ud}^{*})
    &= \frac{g'_8}{f_\pi^2} \left(\frac{2m_K}{\Lambda_\chi}\right)^2 \left( -\frac{1}{3}m_K^2 + \frac{4}{3}m_\pi^2 - \frac{1}{3}p_{\eta_{ud}}^2 - m_\pi^2 Y \right), \label{eq:amp KL to pi pi etaud}\\
    \mathcal{M}(K_L \!\to\! \pi^0 \pi^0 \eta_{s}^{*})
    &= \frac{\sqrt{2}g'_8}{f_\pi^2} \left(\frac{2m_K}{\Lambda_\chi}\right)^2 \left( -\frac{1}{3}m_K^2 + \frac{4}{3} m_\pi^2 - \frac{1}{3} p_{\eta_s}^2 - m_\pi^2 Y \right). \label{eq:amp KL to pi pi etas}
\end{align}
The Dalitz variable $Y$ is defined as
\begin{align}
Y \equiv \frac{s_3 - s_0}{m_\pi^2}, \qquad 
s_i = (p_K - p_i)^2, \qquad 
s_0 = \frac{s_1 + s_2 + s_3}{3},
\end{align}
where $p_1$ and $p_2$ are the four-momenta of the two identical $\pi^0$’s, and $p_3$ corresponds to the four-momentum of the off-shell pseudoscalar $p_{\pi^0}$, $p_{\eta_{ud}}$ or $p_{\eta_s}$, respectively.
The amplitude for $K_L \to \pi^0 \pi^0 a$ can be written by using the mixing angles $\theta_{a\pi}$, $\theta_{a\eta_{ud}}$, and $\theta_{a\eta_s}$, and the amplitudes eqs.~(\ref{eq:amp KL to pi pi pi}--\ref{eq:amp KL to pi pi etas}) as
\begin{align}
\mathcal{M}(K_L \!\to\! \pi^0 \pi^0 a)
&\simeq
\theta_{a\pi}\,\mathcal{M}(K_L \!\to\! \pi^0 \pi^0 \pi^{0*})|_{p_{\pi^0}^2=m_a^2}\notag\\
&\quad+ \theta_{a\eta_{ud}}\,\mathcal{M}(K_L \!\to\! \pi^0 \pi^0 \eta_{ud}^*)|_{p_{\eta_{ud}}^2=m_a^2}\notag\\
&\quad+ \theta_{a\eta_s}\,\mathcal{M}(K_L \!\to\! \pi^0 \pi^0 \eta_{s}^*)|_{p_{\eta_s}^2=m_a^2} \nonumber\\
& = \left( \theta_{a\pi} - \theta_{a\eta_{ud}} - \sqrt{2}\theta_{a\eta_s} \right) \frac{g'_8}{f_\pi^2} \left(\frac{2m_K}{\Lambda_\chi}\right)^2 \left( \frac{1}{3}m_K^2
 - \frac{4}{3}m_\pi^2 + \frac{1}{3}m_a^2 + m_\pi^2 Y \right).
\label{eq:KLpi0pi0a}
\end{align}
The effective interaction term in eq.~\eqref{eq:O8prime} also gives
\begin{align}
    {\cal M}(K_L \to \pi^+ \pi^- \pi^0) = \frac{g'_8}{f_\pi^2} \left(\frac{2m_K}{\Lambda_\chi}\right)^2 \left( \frac{1}{3}m_K^2 - m_\pi^2 + m_\pi^2 Y \right), \label{eq:amp g8prime}
\end{align}
and we can compute the ratio ${\rm Br}(K_L \to \pi^0 \pi^0 a) / {\rm Br}(K_L \to \pi^+ \pi^- \pi^0)$ from these amplitudes. By using ${\rm Br}(K_L \to \pi^+ \pi^- \pi^0) = 0.125$ \cite{ParticleDataGroup:2024cfk} for the normalization, we obtain
\begin{align}
    {\rm Br}(K_L \!\to\! \pi^0 \pi^0 a)
    &\simeq
    f(m_a) \times
    \big(\theta_{a\pi} - \theta_{a\eta_{ud}} - \sqrt{2}\,\theta_{a\eta_s}\big)^2,\label{eq:BR_KLpi0pi0a}
\end{align}
where $f(m_a)$ is a function shown in figure \ref{BrKL}.
At the leading order of chiral Lagrangian, by using the mixing angles in eqs.~\eqref{eq:a pi mixing}, \eqref{eq:a etaud mixing LO}, \eqref{eq:a etas mixing LO}, and the axion mass $m_a$ in eq.~\eqref{eq:axion mass}, the numerical value of the branching ratio of $K_L \to \pi^0 \pi^0 a$ as
\begin{align}
{\rm Br}(K_L \to \pi^0\pi^0 a) |_{\rm (LO)}
\simeq f(m_a) \times 7.9\times 10^{-6} \times \left(\frac{m_a}{\rm 10\,MeV}\right)^2\,.
\label{LOBr}
\end{align}
Note that this numerical value suffers from ${\cal O}(1)$ uncertainty in the mixing angle.
In section~\ref{sec:result}, we will show two analyses: one using the leading-order result \eqref{LOBr} and the other treating $\theta_{a\eta_{ud}}$ and $\theta_{a\eta_s}$ as free parameters in eq.~\eqref{eq:BR_KLpi0pi0a}.

\subsection{Constraints on the axion-electron coupling}
Here we show a summary of experimental constraints on the axion-electron coupling shown in the top panel of figure \ref{fig:summary-plot} 
where the axion mixing angles with $\eta_{ud}$ and $\eta_s$ are evaluated at the leading order in the chiral Lagrangian as in \eqref{eq:a etas mixing LO}. The shaded gray regions in figure~\ref{fig:summary-plot} summarize existing exclusion limits.

The axion can be probed in electron beam-dump experiments
through its coupling to electrons.
In deriving these bounds, the experimental limits on the axion
lifetime reported in each analysis are recast as constraints on
the axion-electron coupling. The relevant limits come from the FNAL E774 experiment~\cite{Bross:1989mp},
the SLAC E141 experiment~\cite{Riordan:1987aw},
the Orsay experiment~\cite{Davier:1986qq},
the NA64 experiment~\cite{Andreev:2021fzd,NA64:2021aiq,Gninenko:2017yus},
and the E137 experiment~\cite{CCM:2021jmk,Bjorken:1988as}.

The axion can be searched for as a resonance at $e^+e^-$ collider experiments, and the competitive bounds are known from the BaBar~\cite{BaBar:2014zli} and KLOE~\cite{Anastasi:2015qla} searches. 
As discussed in section~\ref{Kplus}, we derive the bounds from the $K^+$ decay experiments, NA62~\cite{NA62:2025upx} and E949~\cite{E949:2005qiy}.

The electron anomalous magnetic moment $a_e = (g_e-2)/2$ provides an additional constraint.
The axion couples to electrons through
\begin{align}
    g_a^{e}=\frac{Q_e m_e}{f_a},
\end{align}
and we obtain one-loop contributions as \cite{Dedes:2001nx}
\begin{align}
\Delta a_{e}^{\mathrm{1\text{-}loop}}
   = -\frac{1}{8\pi^2}(g_{a}^{e})^2
      \int_{0}^{1}dx\,
      \frac{(1-x)^{3}}
           {(1-x)^{2} + x(m_{a}/m_{e})^{2}},
\end{align}
which is always negative.
We also obtain two-loop contribution from the Barr--Zee-type two-loop diagrams as \cite{Marciano:2016yhf, Melnikov:2006sr, Stockinger:2006zn, Giudice:2012ms}
\begin{align}
\Delta a_{e}^{\mathrm{2\text{-}loop}}
   = g_{a}^{e}\,g_{a}^{\gamma\gamma}
     \frac{m_{e}}{\pi^{2}}\,
     f_{PS}\!\left[\frac{4\pi f_{\pi}}{m_{a}}\right],
\end{align}
where $g_{a}^{\gamma\gamma}$ is given by eq.~\eqref{gagg}, and 
\begin{align}
f_{PS}[z]
   = \int_{0}^{1}dx\,
     \frac{z^{2}/2}{x(1-x)-z^2}\,
     \log\!\left(\frac{x(1-x)}{z^2}\right).
\end{align}
The two-loop term is proportional to $g_a^{e} g_a^{\gamma\gamma}$ and thus can
be either positive or negative depending on the sign of $Q_e$, whereas the
one-loop term is always negative.
Although $a_e$ has been measured at the level of $1.3 \times 10^{-13}$ \cite{Fan:2022eto}, we note that the structure constant
$\alpha$ entering the calculation of the electron anomalous magnetic
moment suffers from an uncertainty at the level of
${\cal O}(10^{-12})$ because of a discrepancy in the determination of $\alpha$ in different measurements \cite{Parker:2018vye, Morel:2020dww}.
To be conservative, we therefore require that the axion contribution is as small as
\begin{align}
|\Delta a_e| < 10^{-12}.
\end{align}

The blue solid (dashed) curve shows the constraint from the electron anomalous magnetic moment, obtained by requiring $|\Delta a_e|<10^{-12}$ for $Q_e>0$ ($Q_e<0$). 
    The excluded region corresponds to the side of each contour where $|\Delta a_e|$ exceeds $10^{-12}$, as indicated by the direction of the ticks on the contours.

\subsection{Summary of the MeV axion conditions}

In the closing of this section, let us summarize the conditions of the viable MeV axion scenario on top of the conditions (i-iii) given at the beginning of this section:
\begin{itemize}
 \item Pion-phobic: in addition to the choice $Q_u = 2$ and $Q_d = 1$, the significant cancellation to the pion-axion mixing is considered from the higher order, making the SINDRUM bound inconclusive. 
 \item Octet contribution from the higher order, $g_8'\gg g_8$, otherwise, the model is excluded through $K^+ \to \pi^+ a$ with the axion visible decay $a \to e^+ e^-$.
 \item The mass range $2m_e \leq m_a \leq 30~{\rm MeV}$ to kinematically allow $a\to e^+ e^-$ decay and prevent the $B^0 \to K^{*0} e^+ e^-$ bound at LHCb. 
\end{itemize}

\section{\texorpdfstring{$K_L\to\pi^0\pi^0e^+e^-$}{kl to pi0 pi0 e+ e-} at the KTeV experiment}\label{sec:KTeV}
The KTeV experiment at Fermilab has provided the most stringent constraint
on the rare decay $K_L \to \pi^0 \pi^0 e^+ e^-$,
reporting an upper limit of $6.6\times10^{-9}$ at 90\% C.L.~\cite{KTeV:2002tpo}.
This process has the same visible final state as the axion-mediated decay
$K_L \to \pi^0 \pi^0 a$ with $a \to e^+e^-$.
Because the kinematics in the MeV axion model differs from the Standard Model kinematics assumed in the KTeV analysis, a dedicated study is required to assess the applicability of this limit.

In this section, we first review the essential features of the KTeV experiment,
including its detector configuration, beam characteristics,
and the key aspects of its analysis that are relevant to the
$K_L \to \pi^0 \pi^0 e^+ e^-$ search (section~\ref{sec:ktev-overview}).
We then describe the Monte Carlo simulation developed in this work
to reproduce the KTeV setup and evaluate the efficiency
for the axion-induced process $K_L \to \pi^0 \pi^0 a$ with $a \to e^+e^-$ (section~\ref{sec:mc-simulation}).
The obtained efficiency will later be used to reinterpret
the KTeV upper limit in terms of the MeV axion model. Importantly, our fast simulation is validated using the SM decay $K_L\to \pi^0\pi^0e^+ e^-$, and it reproduces the efficiency that is reported by KTeV within a 25\% difference. 

\subsection{Overview of the KTeV Experiment}
\label{sec:ktev-overview}

\begin{figure}[t]
  \centering
  \begin{subfigure}{0.58\linewidth}
    \centering
    \includegraphics[width=\linewidth]{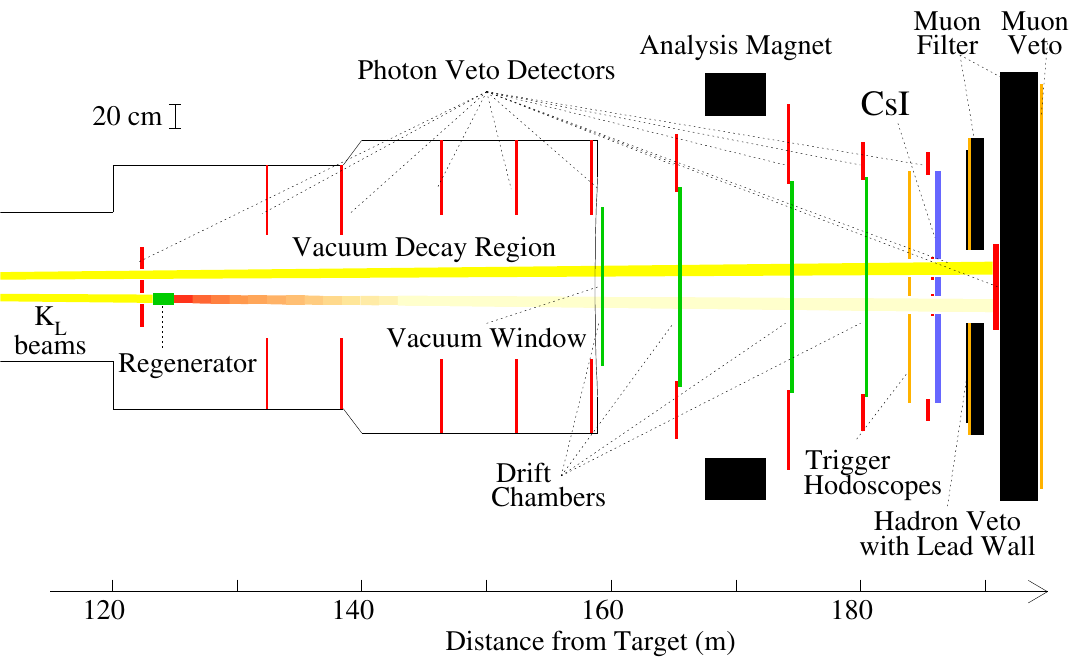}
    \caption{KTeV detector (plan view).}
    \label{fig:ktev-planview}
  \end{subfigure}\hfill
  \begin{subfigure}{0.38\linewidth}
    \centering
    \includegraphics[width=\linewidth]{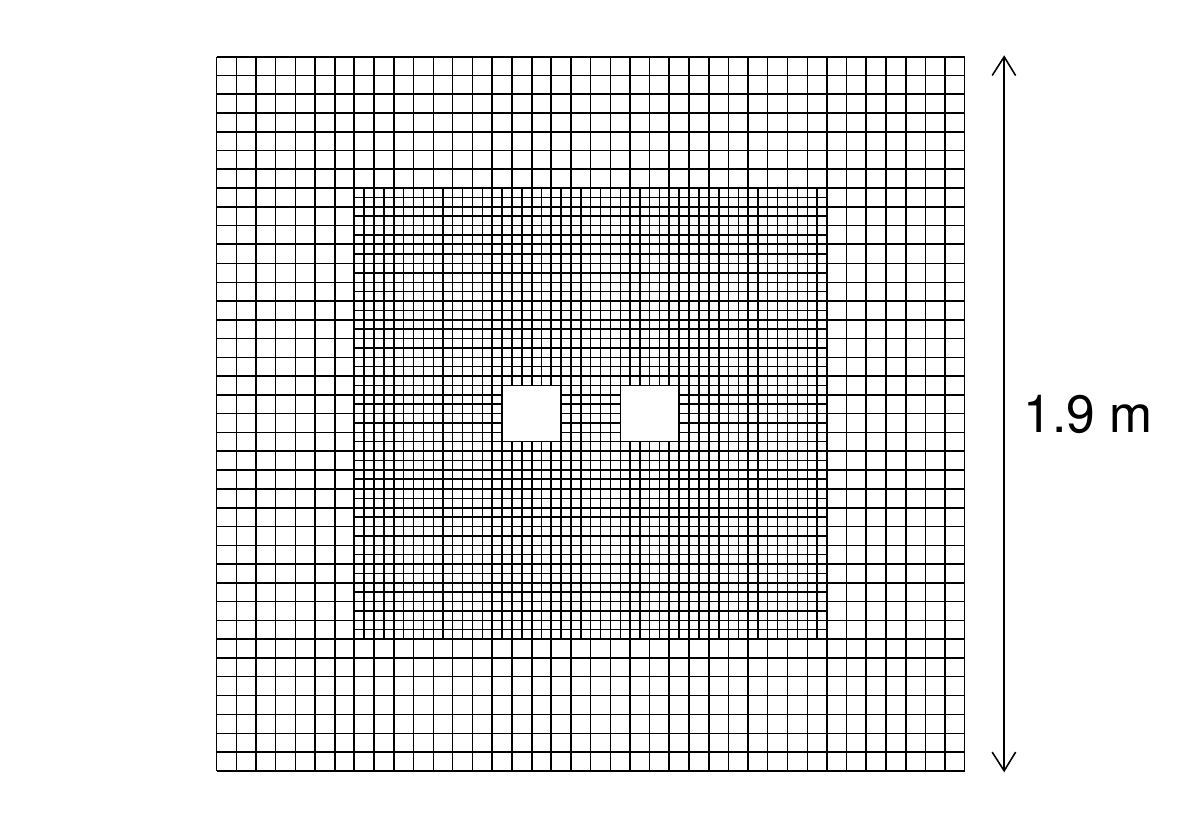}
    \caption{CsI calorimeter front view with two beam holes. }
    \label{fig:csi-schem}
  \end{subfigure}
  \caption{Schematic pictures of the KTeV detector: (a) the $K_L$ beam line with related detectors and (b) the CsI calorimeter view. These figures are taken from ref.~\cite{KTeV:2008nqz}.}
  \label{fig:ktev-figs}
\end{figure}

KTeV is a fixed-target experiment at Fermilab that uses an intense 800~GeV proton beam from the Tevatron to produce two nearly parallel neutral-kaon beams. The detector layout is shown in figure~\ref{fig:ktev-planview} \cite{KTeV:2008nqz}.
Charged-particle tracking is provided by four drift chamber (DC) stations (DC1--DC4), each recording the electron hit positions. An analysis magnet is located between DC2 and DC3, so the electron momentum is obtained from the track deflection between the two DC pairs. At the downstream end, the CsI electromagnetic calorimeter measures photon energy and impact position, and its front face has two square beam holes (figure~\ref{fig:csi-schem}).

\begin{figure}[t]
  \centering
  \includegraphics[width=0.7\linewidth]{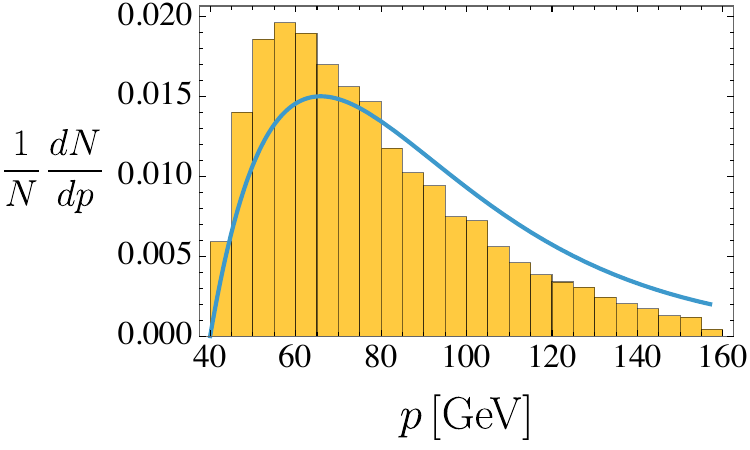}
  \caption{
  The momentum distribution of the incoming $K_L$ flux at the KTeV experiment.
The blue curve corresponds to a smooth fit to the $K_L$ momentum spectrum
adopted in the KTeV analysis~\cite{KTeV:2008nqz}.
The histogram shows the momentum distribution of $K_L$ mesons that decay within
the fiducial region $121~\mathrm{m} \le z \le 158~\mathrm{m}$.
  }
  \label{fig:kl-momentum}
\end{figure}

\begin{figure}[t]
  \centering
  \includegraphics[width=0.7\linewidth]{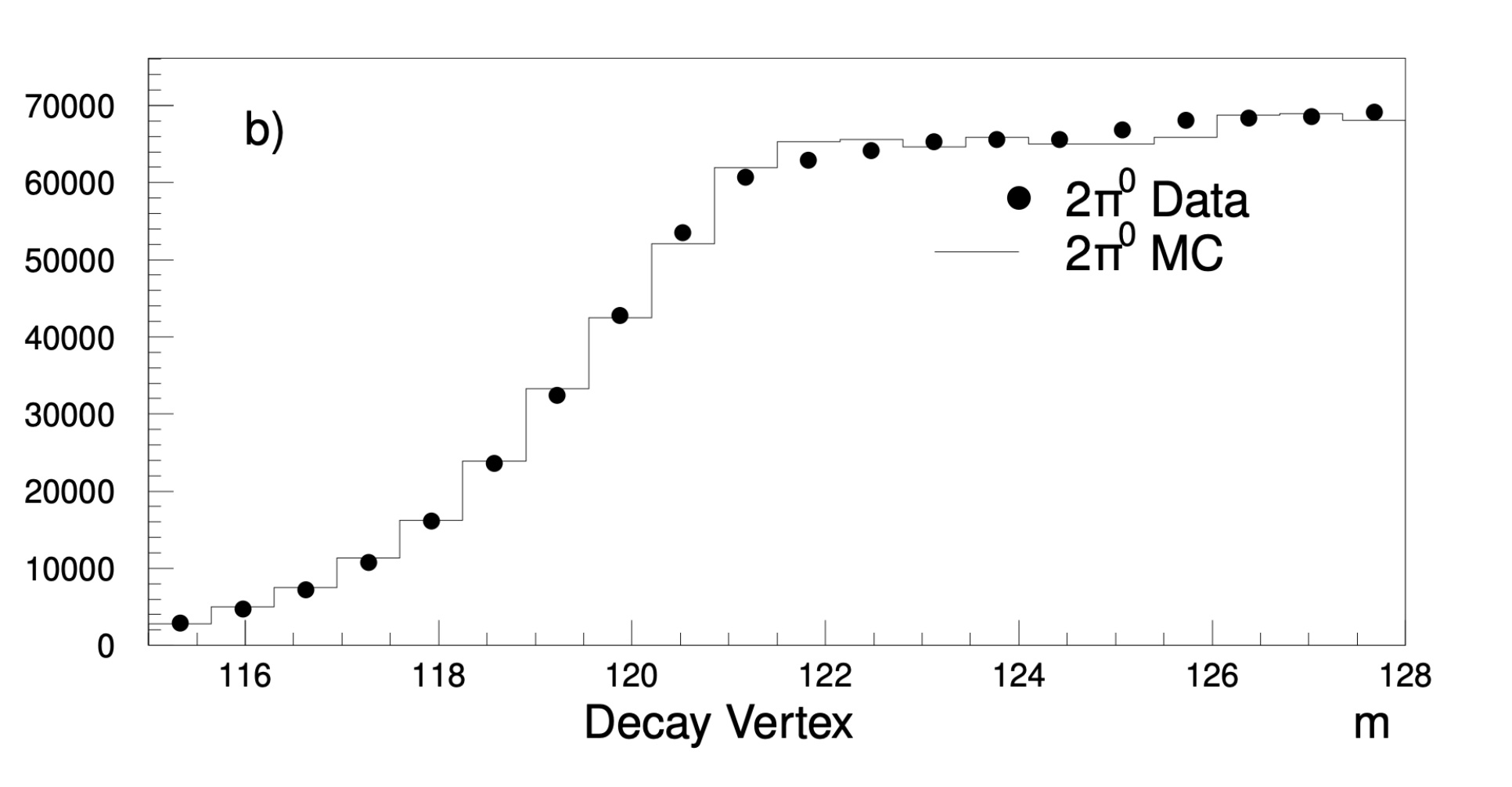}
  \caption{Decay-vertex distribution of reconstructed $K_L$ decays within the fiducial volume $95~\mathrm{m} \le z \le 158~\mathrm{m}$. A sharp turn-on is visible near $z \simeq 121~\mathrm{m}$. This plot is taken from ref.~\cite{KTeV:2008nqz}.}
  \label{fig:z-vertex}
\end{figure}

The momentum distribution of the incoming $K_L$ flux at the KTeV experiment is
shown in figure \ref{fig:kl-momentum}.
The blue curve represents a smooth fit to the $K_L$ momentum spectrum adopted
in the KTeV analysis~\cite{KTeV:2008nqz}.
The histogram shows the momentum distribution of $K_L$ mesons that decay within
the fiducial region $121~\mathrm{m} \le z \le 158~\mathrm{m}$, which is obtained in our simulation.
Here, the coordinate $z$ denotes the distance from the proton-beam target along the beam axis.
In this analysis the fiducial decay volume is defined as $95~\mathrm{m} \le z \le 158~\mathrm{m}$.
The decay-vertex ($z$) distribution reconstructed within this region is displayed in figure~\ref{fig:z-vertex}, which exhibits a sharp turn-on around $z \simeq 121~\mathrm{m}$.


The DC transverse position resolution is
$\sigma^{\mathrm{DC}}=100~\mathrm{\mu m}$, and the CsI calorimeter
position resolution is $\sigma^{\mathrm{CsI}}=1~\mathrm{mm}$ \cite{Bellantoni:2000bj}. The electron-momentum and photon-energy resolutions are parameterized as
\begin{align}
  \left(\frac{\sigma_p}{p}\right)_{e} &= 0.38\% \oplus 0.016\%\,\frac{p}{\rm GeV}, \label{eq:sigmap}\\
   \left(\frac{\sigma_E}{E}\right)_{\gamma}
  &= 0.45\% \oplus \frac{2.0\%}{\sqrt{E/{\rm GeV}}}\,, \label{eq:sigmaE}
\end{align}
  here, ``$\oplus$'' denotes addition in quadrature
($a\oplus b \equiv\sqrt{a^2+b^2}$)~\cite{Bellantoni:2000bj}. These detector parameters are used for the event reconstruction, as discussed in section~\ref{sec:efficiency} and appendix~\ref{Event}.

%%%%%%%%%%%%%%%%%%%%%%%%%%%%%%%%%%%%%%%%%
\subsection{Monte Carlo Simulation of the \texorpdfstring{$K_L \to \pi^0 \pi^0 a$}{kl to pi0 pi0 a} with \texorpdfstring{$a \to e^+e^-$}{a to e+ e-} Signal}
\label{sec:mc-simulation}

To evaluate the acceptance of the KTeV experiment for the axion-induced signal, we developed a dedicated Monte Carlo simulation that closely reproduces the essential features of the KTeV setup. In this simulation and analysis, we consider only the $a\to e^+e^-$ decay mode. As discussed earlier, the reconstructed decay-vertex distribution in KTeV exhibits a sharp turn-on at $z\simeq 121~\mathrm{m}$. For this reason, in the present study, we regard only those $K_L$ decays occurring between $z=121~\mathrm{m}$ and the downstream boundary of the fiducial decay volume at $z=158~\mathrm{m}$ as potential signal events.

The simulation proceeds as follows. We first generate $10^{6}$ incident $K_L$ mesons, with their momenta sampled from the smooth fit to the $K_L$ momentum spectrum
shown in figure~\ref{fig:kl-momentum}, and all $K_L$ mesons are taken to propagate along the beam axis. For each $K_L$, the decay is simulated in its rest frame by generating the Dalitz variables for the three-body process $K_L \to \pi^0 \pi^0 a$. The Dalitz variables are defined as
\[
    s_{12} \equiv (p_{\pi^0_1}+p_{\pi^0_2})^2,\qquad
    s_{23} \equiv (p_{\pi^0_2}+p_{a})^2 .
\]
Within the kinematically allowed region, we sample the Dalitz variables
$(s_{12}, s_{23})$ according to the squared matrix element given in
eq.~\eqref{eq:KLpi0pi0a}. The corresponding energies and three-momenta of the
particles $(\pi^0_1,\pi^0_2,a)$ are then uniquely determined by three-body
kinematics.

After the three-body kinematics are fixed, the two-body decays, such as $a\to e^+e^-$ and $\pi^0\to\gamma\gamma$, are generated isotropically in their respective rest frames. The six final-state particles are then boosted to the laboratory frame according to the sampled $K_L$ momentum. An event is counted as geometrically accepted only if both charged leptons reach the first two drift-chamber stations (DC1 and DC2) and all four photons intersect the CsI calorimeter; events failing to satisfy these conditions are discarded.
In the axion mass range $2~\mathrm{MeV} \le m_a \le 100~\mathrm{MeV}$,
we find that the geometric acceptance is approximately $5.8\%$. 

%%%%%%%%%%%%%%%%%%%%%%%%%%%%%%%%%
\subsection{Efficiency} \label{sec:efficiency}
After evaluating the geometric acceptance, we need to reconstruct $K_L$ decay vertex to perform the event selections. To mimic the vertex reconstruction performed at the KTeV experiment, we incorporate the finite resolutions of the detector components,
in particular, the position and momentum resolutions of the drift chambers and
the energy and position resolutions of the CsI calorimeter. The overall vertex reconstruction procedure is summarized in appendix~\ref{Event}.

With the reconstructed $K_L$ decay vertex, the simulation samples are selected to
suppress background processes, following the KTeV analysis. Although we do not explicitly simulate
background events, this background-rejection procedure closely follows that of
the KTeV experiment, and we impose the same selection cuts used in their
$K_L \to \pi^0\pi^0 e^+e^-$ analysis~\cite{KTeV:2002tpo}.

As the first event selection, we require the
reconstructed invariant mass of the $\pi^0\pi^0 e^+e^-$ system to  lie within
\begin{align}
0.493~{\rm GeV} \le M_{\pi^0\pi^0e^+e^-} \le 0.501~\mathrm{GeV},
\end{align}
and impose a transverse-momentum requirement
\begin{align}
p_x^2 + p_y^2 \le 1.5\times10^{-4}~\mathrm{GeV}^2.
\end{align}
These selections ensure that the reconstructed candidates are compatible with a
$K_L$ decay. The most serious background to the $K_L \to \pi^0\pi^0 e^+e^-$ mode comes from
$K_L \to \pi^0\pi^0\pi^0$ decays.

The first background arises from $K_L \to 3\pi^0$ with a single $\pi^0\to e^+e^-$ decay, which imitates the
signal topology. Following the KTeV analysis,
an event is removed if
\begin{align}
M_{ee} > 0.10~\mathrm{GeV}. \label{eq:meecut}
\end{align}

The second background comes from
$K_L \to 3\pi^0$ with $\pi^0 \to e^+e^- \gamma$ decays, where a Dalitz decay of
one $\pi^0$ produces an $e^+e^-$ pair accompanied by a photon. These events are
removed by discarding candidates with
\begin{align}
M_{ee} < 0.045~\mathrm{GeV}.
\end{align}

The final background arises from events in which a photon converts into an
$e^+e^-$ pair, producing a topology similar to the signal. Following the KTeV
strategy, an event is removed if we find pairs of $e^+\gamma$ and $e^-\gamma$ with
\begin{align}
0.115~{\rm GeV} < M_{e\gamma} < 0.145~\mathrm{GeV},
\end{align}
and the remaining two photons reconstruct a $\pi^0$ within
\begin{align}
|M_{\gamma\gamma} - m_{\pi^0}| < 3~\mathrm{MeV}.
\end{align}
These conditions effectively suppress backgrounds originating from photon
conversions.
\begin{table}[t]
  \centering

  \renewcommand{\arraystretch}{1.5}

  \resizebox{\textwidth}{!}{%
\begin{tabular}{c p{7.4cm} p{6.0cm} r}
    \toprule
    \textbf{Step} &
    \textbf{Selection / Processing} &
    \textbf{Dominant background} &
    \textbf{Event count} \\
    \midrule

    1 &
    Total number of incident $K_L$ &
    -- &
    1,000,000 \\

    2 &
    Number of hit events &
    -- &
    5,812 \\

    3 &
    $
      0.493~{\rm GeV} \le M_{\pi^0\pi^0e^+e^-} \le 0.501~{\rm GeV},\ 
      p_x^2 + p_y^2 \le 1.5\times10^{-4}~{\rm GeV}^2
    $ &
    These selections ensure that the reconstructed candidates originate from the
$K_L$ beam. &
    5,561 \\

    \midrule
    4 &
    Apply cut $M_{ee} > 0.10~\mathrm{GeV}$ &
    $K_L \to 3\pi^0$, $\pi^0 \to e^+e^-$
    (and $\pi^0 \to e^+e^-e^+e^-$) &
    5,561 \\

    5 &
    Apply cut $M_{ee} > 0.045~\mathrm{GeV}$ &
    $K_L \to 3\pi^0$, $\pi^0 \to e^+e^-\gamma$, 1$\gamma$ missing &
    3,119 \\

    6 &
    Reject the event if
    any $e^\pm\gamma\gamma\gamma$ combination satisfies
    $0.115~{\rm GeV}<M_{e\gamma}<0.145~\mathrm{GeV}$ and
    $|M_{\gamma\gamma}-m_{\pi^0}|<3~\mathrm{MeV}$ for the remaining two photons
    &
    $K_L \to 3\pi^0$, $\gamma \to e^+e^-$ (conversion) &
    2,648 \\

    \midrule
    -- & \textbf{Final number of events} & -- & \textbf{2,648} \\
    \bottomrule
  \end{tabular}%
  }
  \caption{Cut flow for the simulated $K_L \to \pi^0 \pi^0 a$ with $a \to e^+e^-$ signal at $m_a = 15\,\mathrm{MeV}$. The event counts show how the signal yield is
reduced by successively applying the same background-rejection cuts as used
in the KTeV $K_L \to \pi^0\pi^0 e^+e^-$ analysis.}
\label{tab:cutflow}
\end{table}
\begin{figure}[t]
  \centering
  \includegraphics[width=0.8\textwidth]{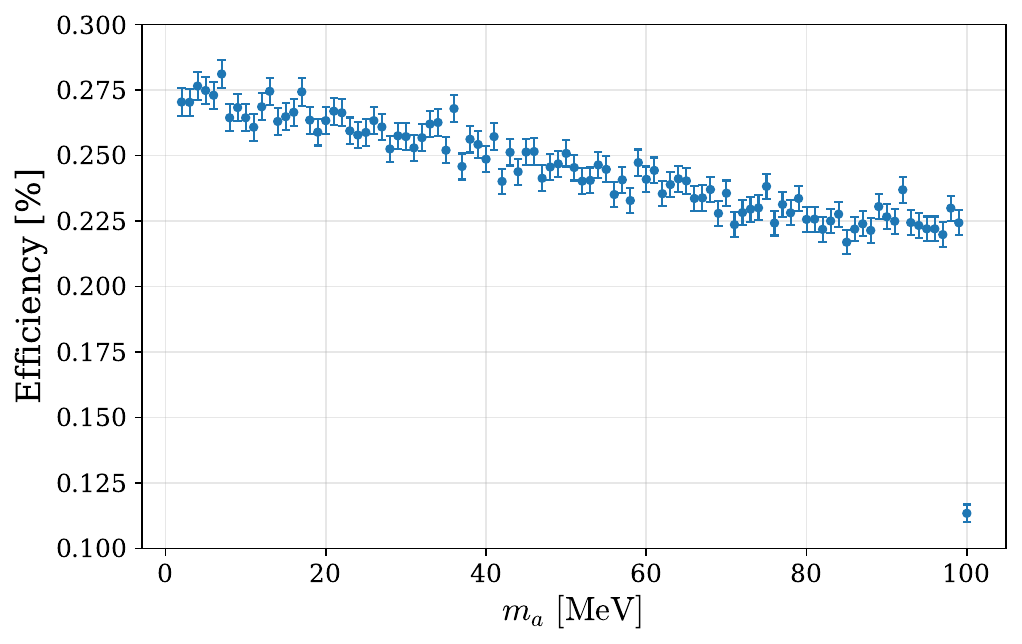}
  \caption{
Event selection efficiency of the $K_L \to \pi^0\pi^0 a$ signal with
$a \to e^+e^-$ as a function of the axion mass $m_a$, assuming prompt axion decay.
The data points represent the efficiencies, and the vertical error
bars indicate the $1\sigma$ statistical uncertainties.
The decrease in efficiency near $m_a \simeq 100~\mathrm{MeV}$ is primarily caused
by the background-rejection cut $M_{ee} > 0.10~\mathrm{GeV}$.
}
  \label{fig:effplo}
\end{figure}

After this background-removal procedure, we obtain the final event selection efficiency. For the benchmark case of $m_a = 15~\mathrm{MeV}$, the cut flow is
summarized in table~\ref{tab:cutflow}, and the resulting efficiency is
approximately $0.26\%$. Figure~\ref{fig:effplo} shows the mass dependence of
the efficiency. In the range $2~\mathrm{MeV} \le m_a < 100~\mathrm{MeV}$, the
efficiency remains nearly constant at the level of $(0.22\!-\!0.27)\%$ for
promptly decaying axions.
The decrease in the efficiency near $m_a \simeq 100~\mathrm{MeV}$ is primarily caused
by the background-rejection cut $M_{ee} > 0.10~\mathrm{GeV}$ in eq.~\eqref{eq:meecut}.

\begin{figure}[t]
  \centering
  \includegraphics[width=0.8\textwidth]{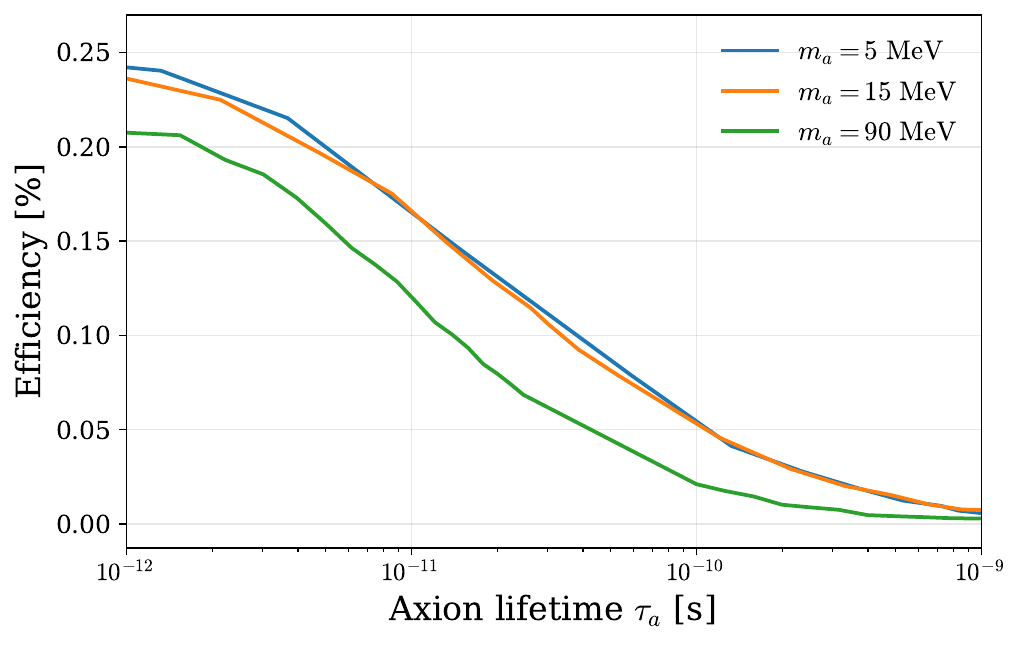}
  \caption{Event selection efficiency as a function of axion lifetime for
$m_a = 5,15,90~\mathrm{MeV}$.
The upper horizontal axis indicates the axion lifetime $\tau_a$,
which is calculated using eq.~\eqref{eq:axion lifetime}.
}
  \label{fig:notpro}
\end{figure}

\begin{figure}[t]
  \centering
  \includegraphics[width=0.8\textwidth]{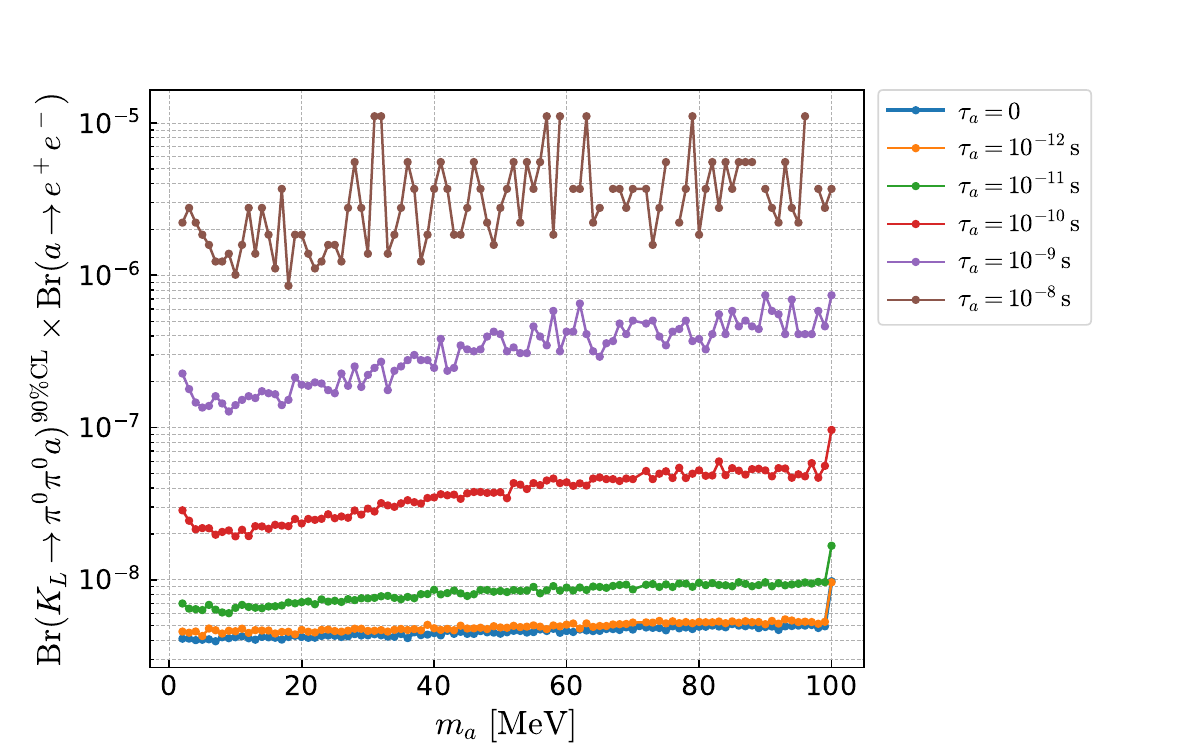}
  \caption{90\% C.L.~bounds on the branching ratio of $K_L\to \pi^0\pi^0 a$ based on eq.~\eqref{eq:boundonpi0-pi0-a}. Relevant axion mass range, $2~{\rm MeV}\leq m_a \leq 100~{\rm MeV}$, and lifetimes are adopted. 
}
  \label{fig:tau}
\end{figure}

We next consider the region in which the axion does not decay promptly. In
parameter regions that evade the  $K^+\to \pi^+ X_{\rm inv}$
bound, the
axion is required to be sufficiently short-lived, and its lifetime is
essentially determined by the electron coupling $Q_e$ as shown in eq.~\eqref{eq:axion lifetime}. To quantify how a
non-prompt decay affects the reconstruction efficiency under this condition, we
evaluate the efficiency as a function of $Q_e$. Figure~\ref{fig:notpro} shows
the result for $m_a = 15~\mathrm{MeV}$. Even for $Q_e \simeq 1/100$, corresponding to the axion lifetime of $\mathcal{O}(10^{-10})$~sec, the
efficiency reduces yet remains of the same order as that in the prompt-decay case, meaning that the KTeV bound is still very effective. 
For the longer lifetime (equivalently smaller $Q_e$), the beam-dump and NA62 ($K^+\to \pi^+ X_{\rm inv}$) bounds appear, and forbid loopholes, as seen in figure~\ref{fig:summary-plot}. Within the axion mass range we consider, $2m_e\leq m_a\leq 30~\rm MeV$, the qualitative behavior stays the same.

To validate our analysis and simulation setup, we also evaluate the event selection efficiency for the
Standard Model process $K_L \to \pi^0\pi^0 e^+e^-$ based on the Sehgal model
with $w_s = 0.9$~\cite{Heiliger:1993qi} (see appendix~\ref{sec:sehgal model}).
The momentum distribution of decay products in the $K_L$ rest frame is simulated according to the matrix element given in eq.~\eqref{eq:Sehgal amplitude}.
This Standard Model signal is fed to our simulation to generate samples, the KTeV reconstruction and event selections are applied as we did in the axion case, and then we obtain an efficiency of $0.168\%$. 
This value is in reasonable agreement with the efficiency of $0.205\%$
reported in the original KTeV analysis~\cite{KTeV:2002tpo}, which was obtained
for the same choice of $w_s$.
This level of agreement indicates that our simulation and analysis 
procedure provide a satisfactory validation of the KTeV analysis.

\subsection{Recasting the KTeV bound \label{sec:bound}}

In the KTeV analysis, the upper bound on the rare SM process with the Sehgal model was obtained as 
\begin{align}
    {\rm Br}(K_L\to \pi^0\pi^0e^+e^-)|_{\rm KTeV}^{90\%\rm CL} = 6.6 \times 10^{-9}
\end{align}
at the 90\% confidence level~\cite{KTeV:2002tpo}. 
The bound on the MeV axion is recast by correcting the efficiencies, 
\begin{align}
    {\rm Br}(K_L\to \pi^0\pi^0a)|^{90\%\rm CL}
    ={\rm Br}(K_L\to \pi^0\pi^0e^+e^-)|_{\rm KTeV}^{90\%\rm CL}
    \cdot \frac{\epsilon_{\rm SM}}{\epsilon_a(m_a, Q_e)}\frac{1}{{\rm Br}(a\rightarrow e^+e^-)},
    \label{eq:boundonpi0-pi0-a}
\end{align}
where $\epsilon_{\rm SM}=0.168\%$ which is based on our simulation. The efficiencies of the MeV axion are at a similar order, as seen in figure~\ref{fig:effplo}, unless the lifetime is longer than $10^{-8}$\,s. We  show the upper bounds on
${\rm Br}(K_L \to \pi^0 \pi^0 a)$ for different lifetimes and masses in figure~\ref{fig:tau}.  
Given that the axion lifetime is less than $10^{-11}$\,s and ${\rm Br}(K_L\to \pi^0\pi^0a)$ is expected to be $\sim 10^{-6}$ in the viable region of parameter space, the KTeV bound on the axions is found to be severe, ruling out the viable window. 

%%%%%%%%%%%%%%%%%%%%%%%%%%%%%%%%%%%%%%%%%%%%%%%%%%%%%%%%%%%%%%
%%%%%%%%%%%%%%%%%%%%%%%%%%%%%%%%%%%%%%%%%%%%%%%%%%%%%%%%%%%%%%
%%%%%%%%%%%%%%%%%%%%%%%%%%%%%%%%%%%%%%%%%%%%%%%%%%%%%%%%%%%%%%

\section{Result}\label{sec:result}
In this section, we present constraints on the MeV axion model derived from the KTeV experiment discussed in the previous section. The branching ratio of $K_L \to \pi^0 \pi^0 a$ is determined by the mixing angles between the axion and neutral mesons, as given in eq.~\eqref{eq:BR_KLpi0pi0a}. In the parameter space of interest, the axion dominantly decays into $e^+ e^-$. To avoid the SINDRUM constraint discussed in section~\ref{sec:axion EFT}, we assume that the mixing with $\pi^0$ is negligible and that the axion mainly mixes with $\eta_{ud}$ and $\eta_s$.
We consider two analyses based on different treatments of the mixing angles $\theta_{a\eta_{ud}}$ and $\theta_{a\eta_s}$: one using the leading order chiral Lagrangian, and the other treating them as free parameters.

\subsubsection*{\texorpdfstring{$\theta_{a\eta_{ud}}$}{theta etaud} and \texorpdfstring{$\theta_{a\eta_s}$}{theta etas} from the leading order chiral Lagrangian}
First, let us use the leading order chiral Lagrangian, and take the mixing angle $\theta_{a\eta_{ud}}$ and $\theta_{a\eta_s}$ given in  eqs.~\eqref{eq:a etaud mixing LO} and \eqref{eq:a etas mixing LO}. Then, ${\rm Br}(K_L \to \pi^0 \pi^0 a)$ is evaluated as a function of $m_a$ as shown in eq.~\eqref{LOBr}, and the experimental bound from the KTeV experiment is evaluated as eq.~\eqref{eq:boundonpi0-pi0-a}.
The result is shown in the bottom panel of 
figure \ref{fig:summary-plot}.
This demonstrates that, at leading order in the chiral Lagrangian, the KTeV
measurement of $K_L \to \pi^0\pi^0 e^+e^-$ imposes an extremely stringent
constraint on this scenario,
though it suffers from the relatively large uncertainty in the mixing angle $\theta_{a\eta_{ud}}$ and $\theta_{a\eta_s}$ from the next-leading order correction in the chiral Lagrangian as discussed in section \ref{sec:axion EFT}.

\subsubsection*{\texorpdfstring{$\theta_{a\eta_{ud}}$}{theta etaud} and \texorpdfstring{$\theta_{a\eta_s}$}{theta etas} as free parameters}
\begin{figure}[t]
  \centering
  \begin{subfigure}{0.42\linewidth}
    \centering
    \includegraphics[width=\linewidth]{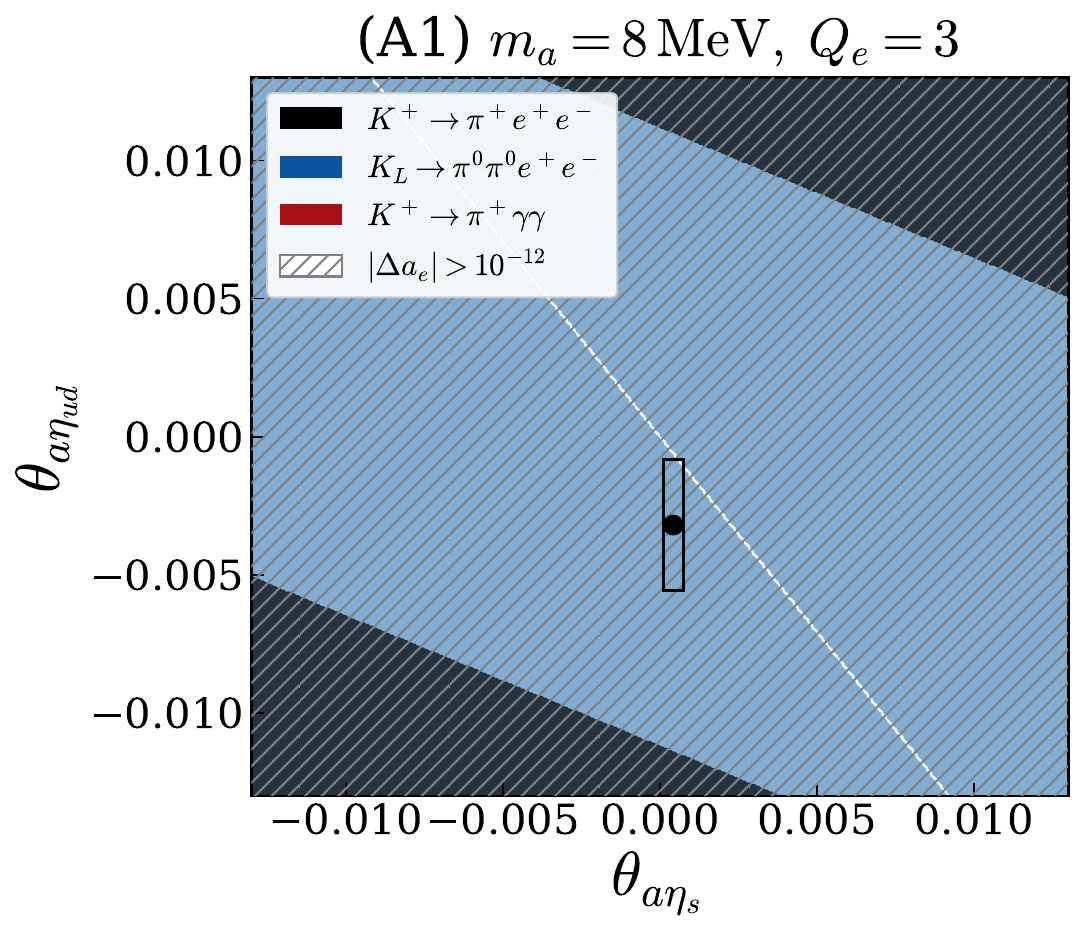}
  \end{subfigure}
  \hfill
  \begin{subfigure}{0.42\linewidth}
    \centering
    \includegraphics[width=\linewidth]{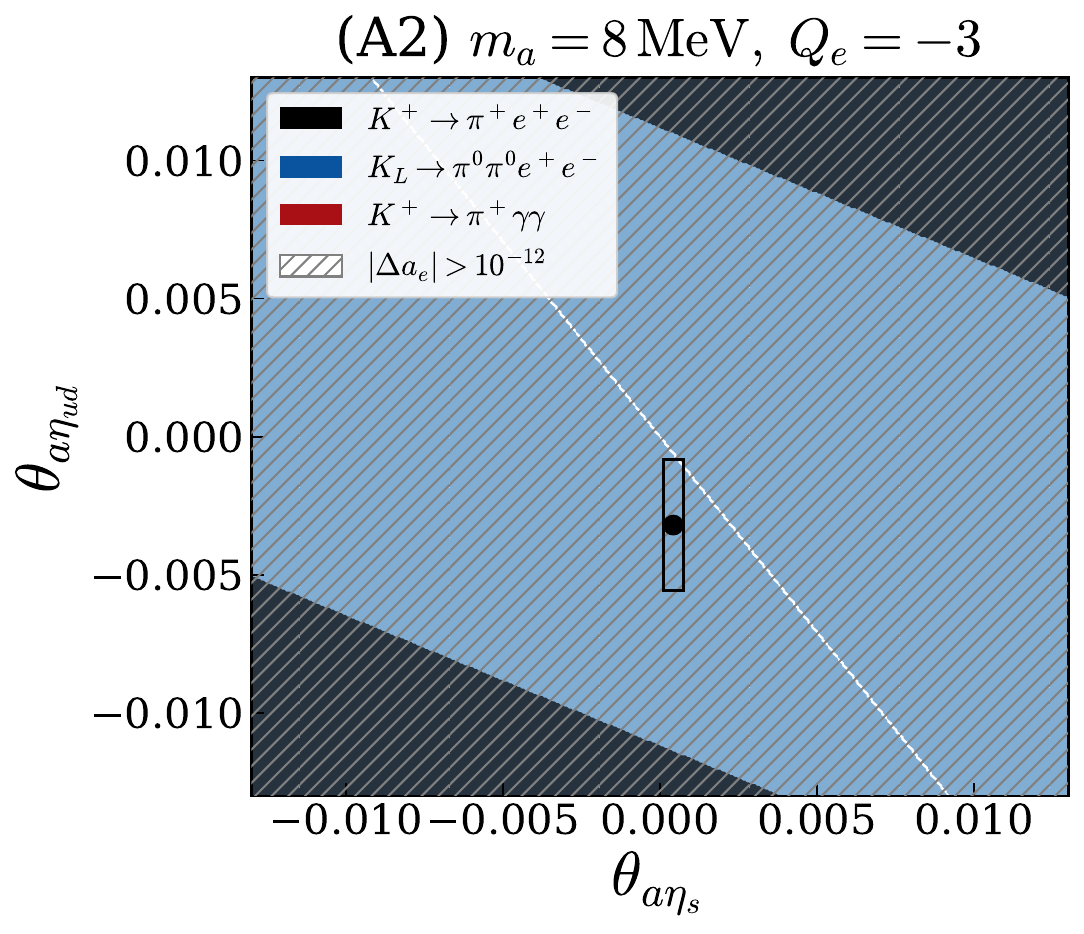}
  \end{subfigure}
~\\[4mm]
  \begin{subfigure}{0.42\linewidth}
    \centering
    \includegraphics[width=\linewidth]{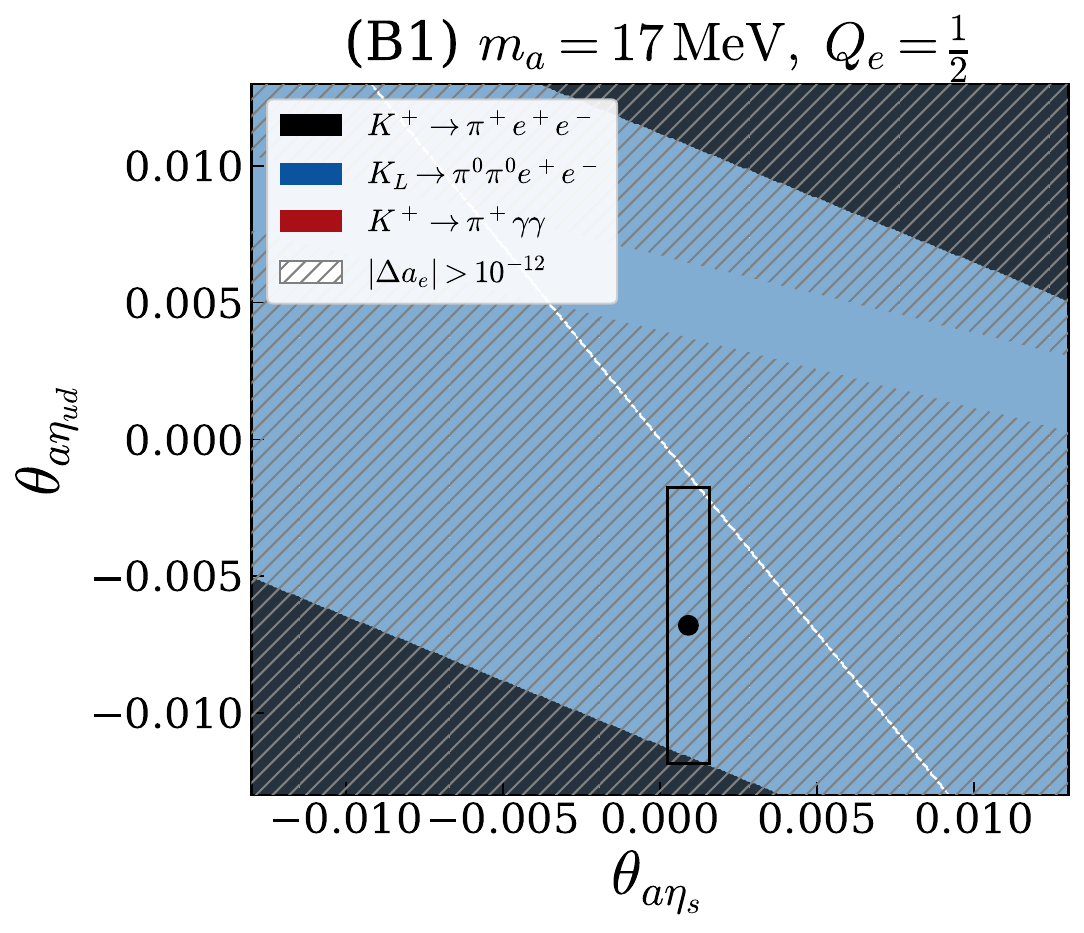}
  \end{subfigure}
  \hfill
  \begin{subfigure}{0.42\linewidth}
    \centering
    \includegraphics[width=\linewidth]{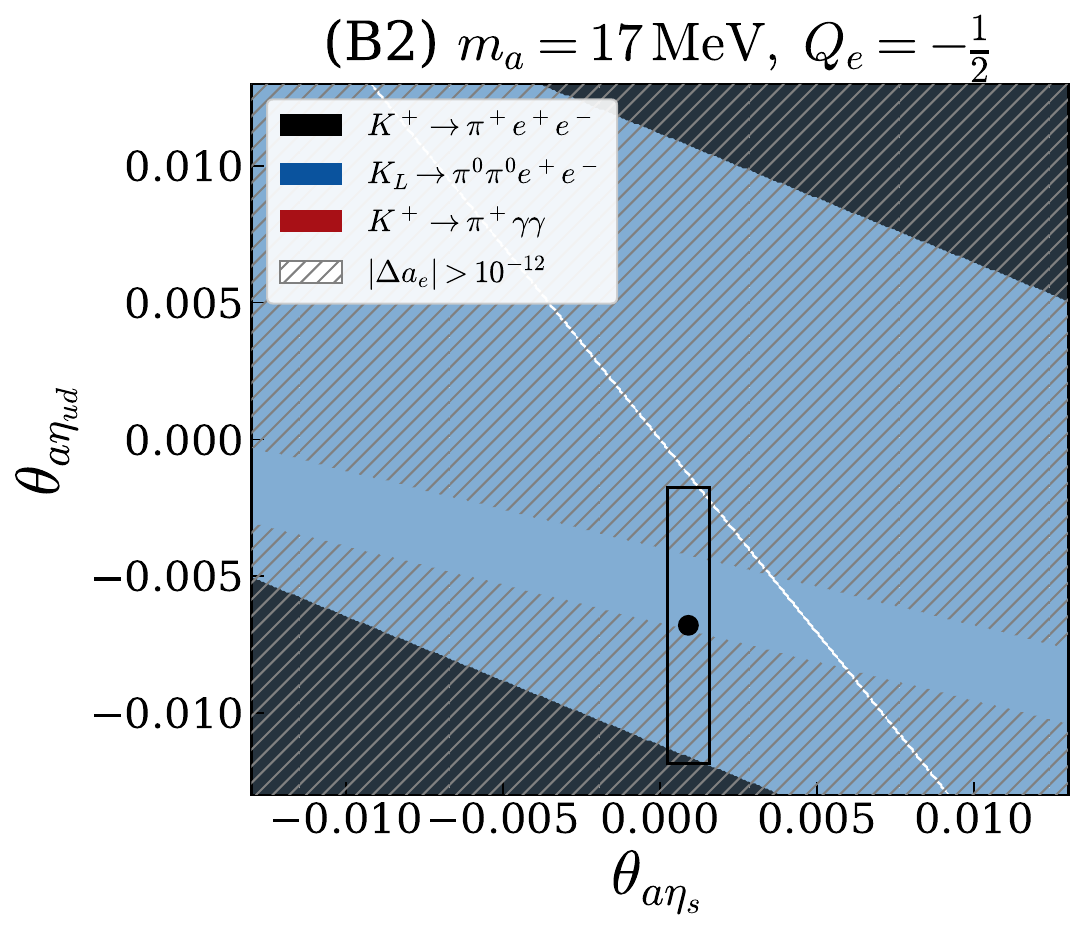}
  \end{subfigure}
~\\[2mm]
  \begin{subfigure}{0.42\linewidth}
    \centering
    \includegraphics[width=\linewidth]{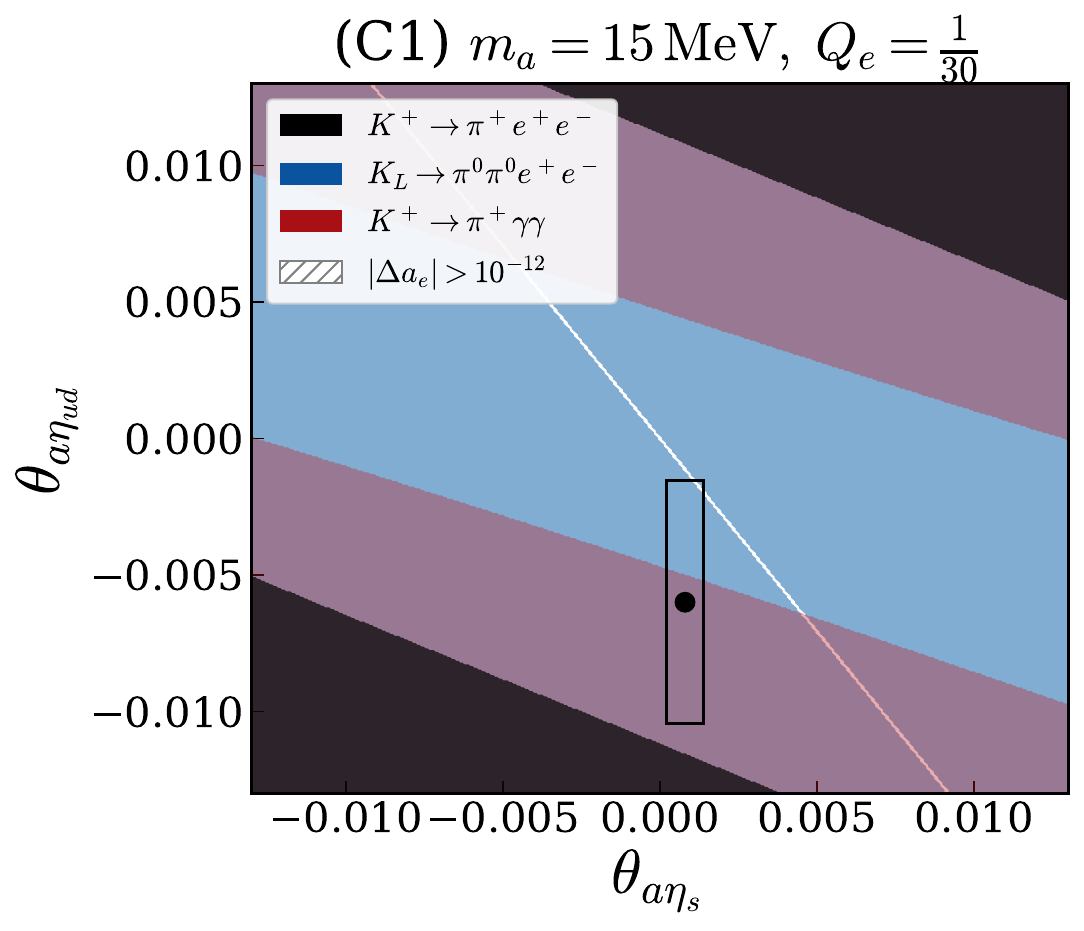}
    \label{15MeVm}
  \end{subfigure}
  \hfill
  \begin{subfigure}{0.42\linewidth}
    \centering
    \includegraphics[width=\linewidth]{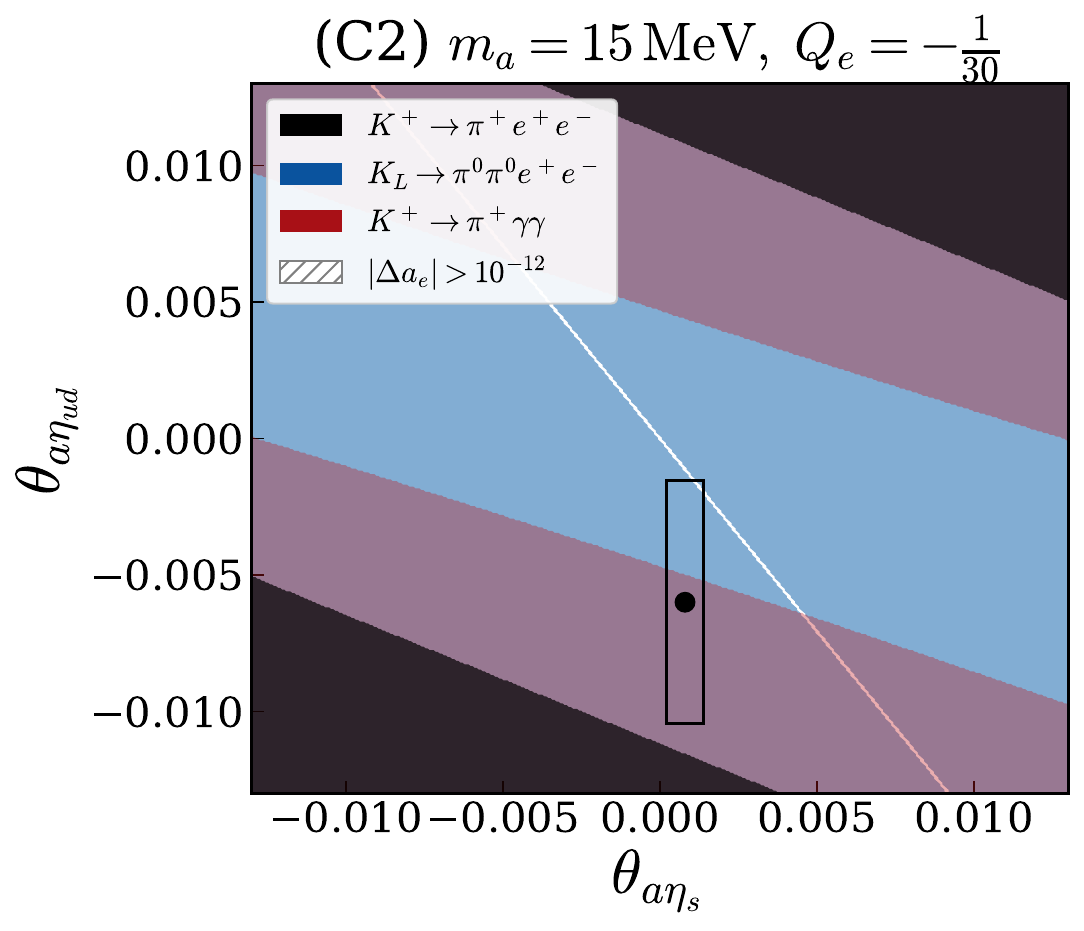}
    \label{15MeVp}
  \end{subfigure}
  \caption{
Constraints on axion-meson mixing angles in the
$(\theta_{a\eta_{ud}},\,\theta_{a\eta_s})$ plane.
Panels (A1) and (A2) correspond to benchmark point A with $(m_a, Q_e)=(8~\mathrm{MeV}, \pm 3)$,
panels (B1) and (B2) to benchmark point B with $(m_a, Q_e)=(17~\mathrm{MeV}, \pm 1/3)$,
and panels (C1) and (C2) to benchmark point C with $(m_a, Q_e)=(15~\mathrm{MeV}, \pm 1/30)$,
as indicated in the top panel of figure~\ref{fig:summary-plot}.
In each row, the left (right) panel corresponds to positive (negative) $Q_e$.
The black dot shows the prediction from the leading order chiral Lagrangian,
and the box indicates the relative uncertainty $2m_K/\Lambda_\chi$ from the
next-leading order corrections.
}\label{fig:thetaud-thetas plane}
\end{figure}

Next, to account for the uncertainty in the mixing angles $\theta_{a\eta_{ud}}$ and $\theta_{a\eta_s}$, we treat them
 as independent
parameters and analyze the constraints directly in the
$(\theta_{a\eta_{ud}},\,\theta_{a\eta_s})$ plane.
For this analysis, we pick up several benchmark points in $(m_a, Q_e/f_a)$-plane as indicated in the bottom panel of figure \ref{fig:summary-plot}, and the results are shown in figure~\ref{fig:thetaud-thetas plane}.
In these plots, we also show the constraint from ${\rm Br}(K^+ \to \pi^+ a) {\rm Br}(a\to e^+e^-) < 10^{-5}$ \cite{Alves:2017avw}
and $|\Delta a_e| < 10^{-12}$. We note that additional constraints from $\eta$ and $\eta^\prime$ decays, discussed in \cite{Alves:2020xhf}, are not shown in the plot, as they lie outside the displayed parameter region.

%%%%%%%%%%%%%%%%%%
%%%%%%%%%%%%%%%%%%

In each panel of figure~\ref{fig:thetaud-thetas plane}, the blue shaded region indicates the exclusion from the KTeV measurement, the black shaded region is ruled out by the $K^{+}$-decay bound,
and the hatched region corresponds to the constraint from the electron
anomalous magnetic moment.
The black dot denotes the point obtained from the leading order chiral
Lagrangian.
The surrounding box visualizes the theoretical uncertainty of the chiral
Lagrangian estimate, which we parameterize by the expansion parameter
$\Delta_\chi =(2m_{K}/\Lambda)^2\approx0.73$ as
\begin{align}
\theta_{a\eta_{ud}}^{\rm (LO)}\left(1-\Delta_\chi\right)    &<\theta_{a\eta_{ud}}<\theta_{a\eta_{ud}}^{\rm (LO)}\left(1+\Delta_\chi\right), \nonumber
\\
\theta_{a\eta_{s}}^{\rm (LO)}\left(1-\Delta_\chi\right)    &<\theta_{a\eta_{s}}<\theta_{a\eta_{s}}^{\rm (LO)}\left(1+\Delta_\chi\right),
\label{eq:uncertainty}
\end{align}
where $\theta_{a\eta_{ud}}^{\rm (LO)}$ and $\theta_{a\eta_s}^{\rm (LO)}$ are in eqs.~\eqref{eq:a etaud mixing LO} and \eqref{eq:a etas mixing LO}, respectively.

We now examine each benchmark scenario in turn, as shown in
figure~\ref{fig:thetaud-thetas plane}.
Panels (A1) and (A2) correspond to benchmark point A with
$(m_a,Q_e)=(8~\mathrm{MeV},\pm 3)$. {The KTeV bound excludes the entire region except for the tiny band that relies on fine cancellations to suppress the production rate by three orders of magnitude from eq.~\eqref{LOBr} (see also eq.~\eqref{eq:BR_KLpi0pi0a}) but barely overlaps the box region. }
In this case, the axion-electron coupling is relatively large, and the constraint from the electron $(g-2)_e$ excludes the parameter space irrespective of the sign of $Q_e$.

Panels (B1) and (B2) correspond to benchmark point B with
$(m_a,Q_e)=(17~\mathrm{MeV},\pm 1/2)$.
For $Q_e>0$, the electron $(g-2)_e$ constraint excludes the entire
box region, leaving no allowed parameter space inside the box.
For $Q_e<0$, an allowed region remains when only the $(g-2)_e$
constraint is considered, but no allowed region survives once
the KTeV constraint is also included.

Finally, panels (C1) and (C2) correspond to benchmark point C with a much smaller
coupling, $(m_a,Q_e)=(15~\mathrm{MeV},\pm 1/30)$.
In this case, the electron $(g-2)_e$ constraint becomes ineffective.
However, the KTeV bound is essentially unchanged and continues to exclude most
of the parameter space, again allowing only a very limited region to remain
viable. For such small values of $Q_e$, the diphoton decay mode becomes
relatively enhanced, and the E949 constraint
 on $K^+\to\pi^+\gamma\gamma$ also becomes relevant.

Taken together, the results shown in figure~\ref{fig:thetaud-thetas plane}
demonstrate that the KTeV constraint provides a stringent bound on this scenario even though the relatively large uncertainties  of the mixing angles $\theta_{a\eta_{ud}}$ and $\theta_{a\eta_{s}}$ are considered. 
All the parameter space is excluded except for the tiny band where uncontrollable higher order corrections in the chiral Lagrangian must be fine-tuned. 
This situation is almost independent of a combination of $(m_a,Q_e)$. 

%%%%%%%%%%%%%%%%%%%%%%%%%%%%%%%%%%%%%%%%%%%%%%%%%%%%%%%%%%%%%%
%%%%%%%%%%%%%%%%%%%%%%%%%%%%%%%%%%%%%%%%%%%%%%%%%%%%%%%%%%%%%%
%%%%%%%%%%%%%%%%%%%%%%%%%%%%%%%%%%%%%%%%%%%%%%%%%%%%%%%%%%%%%%
%%%%%%%%%%%%%%%%%%%%%%%%%%%%%%%%%%%%%%%%%%%%%%%%%%%%%%%%%%%%%%
\section{Conclusion}\label{sec:Conclusion}

The MeV axion model~\cite{Alves:2017avw} is an attractive possibility addressing the strong CP problem at a low new-physics scale. As it can be tested in various experiments, this scenario has received considerable attention. In this paper, we focus on the viable mass range, $2m_e < m_a < 30~{\rm MeV}$, and derive new bounds from kaon experiments. In particular, the constraint from the KTeV search for $K_L \to \pi^0\pi^0 e^+ e^-$, although mentioned in the literature, has not been studied in detail.

Our analysis uses a simulation that accounts for the detector geometry and resolutions, and the setup is validated against the SM result reported by the KTeV experiment. We emphasize that our efficiency results (figures~\ref{fig:effplo} and \ref{fig:notpro}), which lead to the bounds in eq.~\eqref{eq:boundonpi0-pi0-a} as well as figure~\ref{fig:tau}, can be applied to other new particles with masses between 2\,MeV and 100\,MeV.
For the other kaon decays, we obtain the $K^+ \to \pi^+ a$ constraint with $a \to e^+ e^-$ from the E949 measurement, as well as the corresponding constraint from the NA62 measurement in the case where $a$ is effectively invisible. We also include the constraint from the electron $(g-2)_e$ for both signs of $Q_e$.

Based on the leading order chiral Lagrangian and adopting the parameter choice for the octet couplings, $g'_8 \gg g_8$, to avoid the severe bounds on the $K^+ \to \pi^+ a$ process, we find that the KTeV bound excludes the previously viable parameter space, as shown in figure~\ref{fig:summary-plot}.

We further examine potential loopholes arising from higher-order uncertainties in the chiral Lagrangian, as illustrated in figure~\ref{fig:thetaud-thetas plane}. In principle, it is possible to reduce the $K_L \to \pi^0\pi^0 a$ production rate relative to the leading order prediction of eq.~\eqref{LOBr} by three orders of magnitude to evade the KTeV bound, which would require a fine cancellation among the mixing angles given in eq.~\eqref{eq:BR_KLpi0pi0a}. Nevertheless, the corresponding parameter space has little overlap with the theoretical uncertainty bands of eq.~\eqref{eq:uncertainty}. Additionally, the electron $(g-2)_e$ measurement independently constrains the loopholes, except for the small-$|Q_e|$ region where the $K^+ \to \pi^+ a$ limits tend to be stringent.

Therefore, we conclude that the MeV axion model is ruled out --- it may remain viable only if higher-order corrections are unexpectedly large and reduce the production rate by fine-cancellation.
In light of this result, viable axion scenarios would require either a very large decay constant, $f_a \gtrsim 10^8$\,GeV, or an additional mass contribution as in heavy QCD axions.

%%%%%%%%%%%%%%%%%%%%%%%%%%%%%%%%%%%%%%%%%%%%%%%%%%%%%%%%%%%%%%
%%%%%%%%%%%%%%%%%%%%%%%%%%%%%%%%%%%%%%%%%%%%%%%%%%%%%%%%%%%%%%
%%%%%%%%%%%%%%%%%%%%%%%%%%%%%%%%%%%%%%%%%%%%%%%%%%%%%%%%%%%%%%

\section*{Acknowledgements}
The work of RS is supported in part by JSPS KAKENHI Grant Numbers~23K03415, 24H02236, and 24H02244. The work of TI and TY is supported in part by JST SPRING, Grant Number JPMJSP2138. The work of KT is supported in part by the FSU Bridge
Funding 047302. KT thanks the Yukawa Institute for Theoretical Physics at Kyoto University for its hospitality during the completion of this work.

%%%%%%%%%%%%%%%%%%%%%%%%%%%%%%%%%%%%%%%%%%%%%%%%%%%%%%%%%%%%%%
%%%%%%%%%%%%%%%%%%%%%%%%%%%%%%%%%%%%%%%%%%%%%%%%%%%%%%%%%%%%%%
%%%%%%%%%%%%%%%%%%%%%%%%%%%%%%%%%%%%%%%%%%%%%%%%%%%%%%%%%%%%%%

\appendix
\section{Note on the Sehgal model for \texorpdfstring{$K_L \to \pi^0 \pi^0 e^+ e^-$}{kl to pi0 pi0 e+ e-}}\label{sec:sehgal model}
The KTeV experiment  \cite{KTeV:2002tpo} estimated the signal efficiency for the Standard Model contribution based on the Sehgal model \cite{Heiliger:1993qi}. In this appendix, we summarize the results in ref.~\cite{Heiliger:1993qi}, and also correct some typos in the original paper.

\subsection{The matrix element and parameters}
First we introduce the transition amplitude ${\cal A}$ for $K_L \to \pi^0 \pi^0 e^+ e^-$ and summarize the parameters in the amplitude.

The transition amplitude is given in eq.~(10) of ref.~\cite{Heiliger:1993qi}.
Here we rewrite the amplitude using the definition of the couplings $g_{M1}$ and $g_{E2}$ in ref.~\cite{Sehgal:1992wm} to use the numerical values in ref.~\cite{Sehgal:1992wm}
because the numerical value of $g_{M1}$ and $g_{E2}$ used in ref.~\cite{Heiliger:1993qi} is not clearly written in ref.~\cite{Heiliger:1993qi} itself. 
We should be careful about the difference in the definition of $g_{M1}$ and $g_{E2}$ in ref.~\cite{Heiliger:1993qi} and ref.~\cite{Sehgal:1992wm}.
For this purpose, we compare the transition amplitude for $K_L \to \pi^+ \pi^- e^+ e^-$ given in eq.~(4) in ref.~\cite{Heiliger:1993qi} and eq.~(7) in ref.~\cite{Sehgal:1992wm}.
$g_{M1}$ and $g_{E2}$ in ref.~\cite{Sehgal:1992wm} are dimensionless parameters.
By comparing the amplitudes for $K_L \to \pi^+ \pi^- e^+ e^-$, we should understand $g_{M1}$ in ref.~\cite{Heiliger:1993qi} has mass dimension 1 and should be equivalent to $e f_S g_{M1}$ in ref.~\cite{Sehgal:1992wm}. Similarly, $g_{E2}$ in ref.~\cite{Heiliger:1993qi} should have mass dimension 1 and should be equivalent to $e f_S g_{E2}$ in ref.~\cite{Sehgal:1992wm}.

Applying the definition of $g_{E2}$ in ref.~\cite{Sehgal:1992wm},  we rewrite the matrix element for $K_L \to \pi^0(p_1) \pi^0(p_2) e^+(k_+) e^-(k_-)$ given in eq.~(10) in ref.~\cite{Heiliger:1993qi} as
\begin{align}
    {\cal A}(K_L \to \pi^0 \pi^0 e^+ e^-) 
    =&
    \biggl[
        e^2 f_S
        \frac{g_{E2}}{m_K^4} \frac{(p_1-p_2)\cdot k}{\Lambda^2} (k\cdot p_2 p_1^\mu - k\cdot p_1 p_2^\mu ) \frac{1}{k^2} \nonumber\\
    & \quad
        - \frac{e^2 f_S}{3\sqrt{2}} \frac{\langle R^2 \rangle}{(m_K^2 - m_\pi^2)}  \left( \frac{s_\pi - m_\pi^2}{s_\pi - m_K^2} + \frac{3w_s-1}{\log(m_K^2 / m_\pi^2)} \right) p^\mu \biggr] \nonumber\\
    & \qquad\times \bar u (k_-) \gamma_\mu v(k_+). \label{eq:Sehgal amplitude}
\end{align}
Also we interpret the definition of $\delta_0$ and $\delta_2$ to be consistent with the sign $\pm$ in $H_1$ of ref.~\cite{Heiliger:1993qi} as
\begin{align}
    g_{E2} = \pm |g_{E2}| e^{i\delta_2}, \qquad
    f_S = |f_S| e^{i\delta_0}. \label{eq:gE2 and fS}
\end{align}

The amplitude in eq.~\eqref{eq:Sehgal amplitude} has five parameters; $f_S$, $g_{E2}$, $\Lambda$, $\langle R^2\rangle$, and $w_s$.
$f_S$ is calculated from the definition in eq.~(11) in ref.~\cite{Heiliger:1993qi} and $\Gamma(K_S \to \pi^+ \pi^-)$ as
\begin{align}
    |f_S| = 3.92 \times 10^{-4}~{\rm MeV}.
\end{align}
Here we have used $\tau_{K_S} = 8.954 \times 10^{-11} \text{ s}$, ${\rm Br}(K_S \to \pi^+ \pi^-) = 0.692$, $m_K = 497.6~{\rm MeV}$, and $m_{\pi^\pm} = 139.6~{\rm MeV}$ \cite{ParticleDataGroup:2024cfk}.
$g_{E2}$ and $\Lambda^2$ appear as $g_{E2}/\Lambda^2$ in the amplitude, and it can be obtained from eqs.~(5, 6) in ref.~\cite{Heiliger:1993qi} as $g_{E2} / \Lambda^2 = g_{M1} \times 1.68 \times 10^{-6}~{\rm MeV}^{-2}$.
By using $g_{M1} = 0.76$ \cite{Sehgal:1992wm}, we obtain
\begin{align}
    \frac{g_{E2}}{\Lambda^2} = 1.28 \times 10^{-6}~{\rm MeV}^{-2}.
\end{align}
We also take the mean square charge radius of $K^0$ as \cite{Heiliger:1993qi}
\begin{align}
    \langle R^2 \rangle = -0.07~{\rm fm}^2.
\end{align}
For $w_s$, we take $w_s = 0.5$ in this appendix to check the consistency with ref.~\cite{Heiliger:1993qi}.
However, in section \ref{sec:efficiency}, we take $w_s = 0.9$ for the comparison with the efficiency reported by the KTeV experiment \cite{KTeV:2002tpo}.

\subsection{Partial decay width and invariant mass distributions}
Let us calculate the partial decay width $d^2 \Gamma / ds_\pi ds_\ell$ where $s_\pi = (p_1 + p_2)^2$ and $s_\ell = (k_+ + k_-)^2$, and discuss the distributions of $s_\pi$ and $s_\ell$.
$d^2 \Gamma / ds_\pi ds_\ell$ is obtained from the following integral:
\begin{align}
    \frac{d^2\Gamma}{d s_\pi ds_\ell}
    =& \frac{1}{16\pi^2 m_K}
        \left( \frac{1}{8\pi} \frac{2P}{m_K} \right) \nonumber\\
    & \times
        \left( \frac{1}{8\pi} \frac{2p_\pi^*}{\sqrt{s_\pi}} \frac{1}{2}\int_0^\pi d\theta_\pi^*  \sin\theta_\pi^* \right)
        \left( \frac{1}{8\pi} \frac{2p_\ell^*}{\sqrt{s_\ell}} \frac{1}{4\pi} \int_0^{2\pi} d\phi_\ell^* \int_0^\pi d\theta_\ell^* \sin\theta_\ell^* \right)
        |{\cal A}|^2. \label{eq:dgammadsds}
\end{align}
where $P$ is the momentum of $\pi^0\pi^0$ system (or $e^+e^-$ system) in the $K_L$ rest frame, which is given by 
\begin{align}
    P = \frac{1}{2m_K} \sqrt{(m_K^2 - s_\pi - s_\ell)^2 - 4s_\pi s_\ell}.
\end{align}
$p_\pi^*$ and $p_\ell^*$ are the momenta of $\pi^0$ in $\pi^0 \pi^0$ rest frame and the momentum of $e$ in $e^+e^-$ rest frame, respectively:
\begin{align}
    p_\pi^* = \frac{1}{2}\sqrt{s_\pi - 4m_\pi^2}, \qquad
    p_\ell^* = \frac{1}{2}\sqrt{s_\ell - 4m_\ell^2}.
\end{align}

By plugging eq.~(\ref{eq:Sehgal amplitude}) into eq.~(\ref{eq:dgammadsds}), we obtain
\begin{align}
    \frac{d\Gamma}{d s_\pi ds_\ell}
    =
    \frac{1}{2^9 \pi^5 m_K^3} \left( 1 - \frac{4 m_\ell^2}{s_\ell} \right)^{
        1/2 % 3/2 in Heiliger:1993qi
    } \left( 1 - \frac{4 m_\pi^2}{s_\pi} \right)^{1/2}
    X \left(I_1 - \frac{1}{3}I_2\right),
\end{align}
where $I_1$, $I_2$, and $X$ are defined as
\begin{align}
    I_1 =& \frac{1}{4} \left[ \left( 1 + \frac{4m_\ell^2}{s_\ell}\right) H_1 + \frac{3}{2} \left( 1 + \frac{4m_\ell^2}{3s_\ell }\right) H_2 \right], \\
    I_2 =& -\frac{1}{4} \left( 1 - \frac{4m_\ell^2}{s_\ell} \right) \left( H_1 - \frac{1}{2} H_2 \right), \\
    X =& (s^2 - s_\pi s_\ell)^{1/2}.
\end{align}
$H_1$, $H_2$, and $s$ in the above expressions are defined as
\begin{align}
    H_1 =& X^2 \Biggl[
        \frac{e^2}{40} \frac{|g_{E2}|^2}{m_K^8} \frac{\sigma_\pi^2}{\Lambda^4 s_\ell^2}(
            \sigma_\pi^2 X^4 + \sigma_\pi^2 s^4 
        ) \nonumber\\
    & \quad
        + \frac{e^4}{36} |f_S|^2 \frac{\langle R^2 \rangle^2}{(m_K^2 - m_\pi^2)^2} 
        \left( \frac{s_\pi - m_\pi^2}{s_\pi - m_K^2} + \frac{3w_s-1}{\log(m_K^2 / m_\pi^2)} \right)^2 \nonumber\\
    & \quad
        \pm \frac{\sqrt{2} e^3}{36} \frac{|g_{E2}|}{m_K^4} \frac{|f_S|}{\Lambda^2} 
        \frac{
            \sigma_\pi^2 
        X^2}{s_\ell} \frac{\langle R^2 \rangle}{(m_K^2 - m_\pi^2)} \left( \frac{s_\pi - m_\pi^2}{s_\pi - m_K^2} + \frac{3w_s-1}{\log(m_K^2 / m_\pi^2)} \right) \cos(\delta_2 - \delta_0) \Biggr] \nonumber\\
    &
    + 2Xs\sigma_\pi \Biggl[
        -\frac{e^2}{40} \frac{|g_{E2}|^2}{m_K^8} \frac{\sigma_\pi^3 X^3}{\Lambda^4}  \frac{s}{s_\ell^2} \nonumber\\
    & \quad
        \mp \frac{\sqrt{2} e^3}{72} \frac{|g_{E2}|}{m_K^4} \frac{|f_S|}{\Lambda^2} \frac{\langle R^2 \rangle}{(m_K^2 - m_\pi^2)} \left( \frac{s_\pi - m_\pi^2}{s_\pi - m_K^2} + \frac{3w_s-1}{\log(m_K^2 / m_\pi^2)} \right) \sigma_\pi X \frac{s}{s_\ell}\cos(\delta_2 - \delta_0) \Biggr], \\
    H_2 =& \sigma_\pi^4 s_\pi s_\ell \frac{e^2}{60} \frac{|g_{E2}|^2}{m_K^8} \frac{X^2}{\Lambda^4}  \frac{s^2}{s_\ell^2}, \\
    s =& \frac{1}{2} (m_K^2 - s_\pi - s_\ell).
\end{align}
where
\begin{align}
    \sigma_\pi = \left( 1 - \frac{4m_\pi^2}{s_\pi}\right)^{1/2}.
\end{align}
From this result, we found several typos in eq.~(12) of  ref.~\cite{Heiliger:1993qi};
the power of $(1 - 4m_\ell^2/s_\ell)$ in eq.~(12) should be $1/2$ instead of $3/2$, $(X^4 + \sigma_\pi^2 s^4)$ in the first term in the parenthesis of the definition of $H_1$ should be $(\sigma_\pi^2 X^4 + \sigma_\pi^2 s^4)$, and $\sigma_\pi$ in the third term in the parenthesis of the definition of $H_1$ should be $\sigma_\pi^2$.

\begin{figure}
\centering
\includegraphics[width=0.49\hsize]{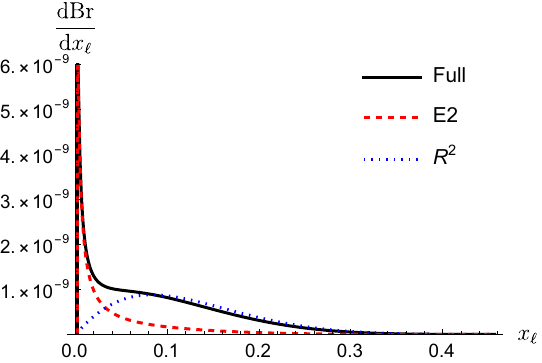}
\includegraphics[width=0.49\hsize]{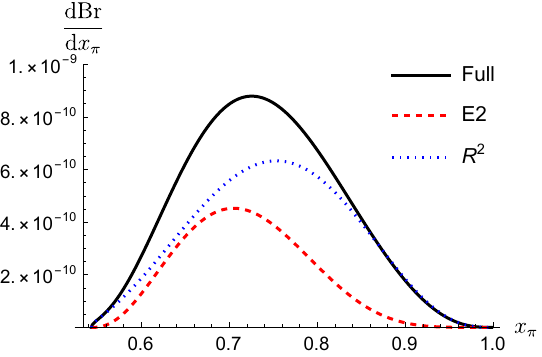}
\caption{The invariant mass distribution of $K_L \to \pi^0 \pi^0 e^+ e^-$. The left panel shows $d{\rm Br}/dx_\ell$ where $x_\ell \equiv \sqrt{s_\ell}/m_K$, and the right panel shows $d{\rm Br}/dx_\pi$ where $x_\pi \equiv \sqrt{s_\pi}/m_K$. In each panel, the black curve shows the distribution from the full amplitude eq.~\eqref{eq:Sehgal amplitude}, the red curve shows the contribution from $g_{E2}$ term, and the blue curve shows the contribution from $\langle R^2\rangle$ term. For the black curve, we take a positive sign in eq.~\eqref{eq:gE2 and fS} with $\delta_0 = \delta_2$.}
\label{fig:sehgal distribution}
\end{figure}

Figure \ref{fig:sehgal distribution} shows the invariant mass distribution of $e^+ e^-$ and $\pi^0 \pi^0$. We can see the invariant mass of $e^+ e^-$ has a sharp peak around the kinematical threshold. 
The origin of this behavior is the off-shell $\gamma^*$ contribution in $g_{E2}$ term as indicated in eq.~\eqref{eq:Sehgal amplitude}. We can see that figure \ref{fig:sehgal distribution} is consistent with figures 3 and 4 of ref.~\cite{Heiliger:1993qi} at the level of ${\cal O}(10)\%$.

\subsection{Integrated branching fraction}
Finally, let us calculate the integrated branching fraction. We perform the integration of $s_\pi$ and $s_\ell$ in the range of $4m_\pi^2 \leq s_\pi \leq (m_K - 2m_e)^2$ and $4m_e^2 \leq s_\ell \leq (m_K - \sqrt{s_\pi})^2$.
Then, we obtain the integrated branching fraction as
\begin{align}
    {\rm Br}(K_L \to \pi^0 \pi^0 e^+ e^-)|_{\rm total} &= (1.96,~2.68) \times 10^{-10}. \label{eq:Br3}
\end{align}
Two values in the above expression correspond to the two possible signs in eq.~(\ref{eq:gE2 and fS}) with $\delta_2 = \delta_0$.
To compare with the numerical value reported in ref.~\cite{Heiliger:1993qi}, let us pick up the $g_{E2}$ term in eq.~(\ref{eq:Sehgal amplitude}), then, we obtain
\begin{align}
    {\rm Br}(K_L \to \pi^0 \pi^0 e^+ e^-)|_{E2} &= 8.30 \times 10^{-11}. \label{eq:Br1}
\end{align}
On the other hand, if we only pick up the $\langle R^2 \rangle$ term in eq.~(\ref{eq:Sehgal amplitude}), we obtain
\begin{align}
    {\rm Br}(K_L \to \pi^0 \pi^0 e^+ e^-)|_{\langle R^2 \rangle} &= 1.49 \times 10^{-10}. \label{eq:Br2}
\end{align}
We find eqs.~(\ref{eq:Br3}), (\ref{eq:Br1}), and (\ref{eq:Br2}) are consistent with eq.~(13) in ref.~\cite{Heiliger:1993qi} in $10$--$40$\%.

\section{Event reconstruction}
\label{Event}
In the KTeV experiment, photon momenta are not measured directly, and their
directions must be reconstructed using calorimetric information together with
the charged-particle tracking provided by the drift chambers.
This reconstruction requires a determination of the $K_L$ decay vertex, making
the vertex reconstruction an essential ingredient of the analysis.

The directly measured observables consist of the hit
positions of charged particles in the drift chambers and the energies and
hit positions of photons in the CsI calorimeter.
For the electron and positron, the tracking information provides the momentum
direction while its magnitude is obtained from the track curvature
due to the magnet.
For the photons, only the energy and the transverse position at the calorimeter
front face are measured.
The decay vertex of $K_L$  is reconstructed by  the charged-particle tracking information from the drift chambers under the assumption that the $K_L$ travels along the nominal beam line.
Additionally, we can reconstruct the vertex by the photon information because each photon pair provides an independent estimate of the decay vertex through the $\pi^0$ mass constraint.

To obtain the best vertex position of the $K_L $ decay  with the different reconstruction methods, we minimize the global $\chi^2$ function, 
\begin{align}
\chi^2_{\rm tot} (z_{\rm v})= \chi_{\rm track}^2(z_{\rm v}) + \chi_{\rm cal}^2(z_{\rm v}),
\end{align}
 for the every possible photon pairing where $z_{\rm v}$ is the vertex position along the $K_L$ beam line. As we will discuss in the following, this $\chi^2$ accounts for the measurement uncertainties which are described in section~\ref{sec:ktev-overview}.  
 The correct photon pairing is also inferred from the $\chi^2_{\rm tot}$ fitting.

\subsection{\texorpdfstring{$\chi_{\rm track}^2$}{chi track squareds}}

The tracking contribution $\chi_{\rm track}^2$ quantifies the consistency of the
electron and positron trajectories measured by the first two drift chambers
with a common decay vertex.
Each charged particle is assumed to follow a straight-line trajectory between
the two drift chambers, and the corresponding $\chi^2$ is written as \cite{Senyo:1999pi}
\begin{align}
\chi_{\rm track}^2(z_{\rm v})
= \sum_{i=e^+,e^-}\sum_{\mathrm{DC}=1,2}
\left[
\frac{(x_{\mathrm{DC}}^i - s_x^i (z_{\mathrm{DC}}^i - z_{\rm v}))^2}{\sigma_x^2}
+
\frac{(y_{\mathrm{DC}}^i - s_y^i (z_{\mathrm{DC}}^i - z_{\rm v}))^2}{\sigma_y^2}
\right].
\end{align}
Here the position resolution is given by $\sigma_x=\sigma_y=100~\mathrm{\mu m}/\sqrt{2}$, and $x_{\mathrm{DC}}^i$, $y_{\mathrm{DC}}^i$ and $z_{\mathrm{DC}}^i$ are observed (simulated) quantities. 
While the track slopes, $s_x^i$ and $s_y^i$, are free parameters, they are
determined analytically by minimizing $\chi_{\rm track}^2$ for a fixed vertex position $z_{\rm v}$ as 
\begin{align}
    s_x^i = \frac{\sum_{\mathrm{DC}=1,2}(z_{\mathrm{DC}}^i - z_{\rm v}) x_{\mathrm{DC}}^i}
    {\sum_{\mathrm{DC}=1,2}(z_{\mathrm{DC}}^i - z_{\rm v})^2}, 
\quad 
    s_y^i = \frac{\sum_{\mathrm{DC}=1,2}(z_{\mathrm{DC}}^i - z_{\rm v}) y_{\mathrm{DC}}^i}
    {\sum_{\mathrm{DC}=1,2}(z_{\mathrm{DC}}^i - z_{\rm v})^2}
\end{align}
the track slopes $s_x^i$ and $s_y^i$ are
determined analytically by minimizing $\chi_{\rm track}^2$.
This track $\chi^2$ typically has a greater sensitivity to the vertex position than $\chi^2_{\rm cal}$.

\subsection{\texorpdfstring{$\chi_{\rm cal}^2$}{chi cal squared}}

The calorimeter contribution $\chi_{\rm cal}^2$ is constructed using the
kinematic constraint from $\pi^0\to\gamma\gamma$ decays.
For a given photon pair $(\gamma_1,\gamma_2)$, denoted by the $A$ pair, the decay vertex position inferred
from the $\pi^0$ mass constraint is given by
\[
z_A = z_{\rm cal} - \frac{\sqrt{E_1E_2}}{m_\pi}\,r_{12},
\]
where $E_{1,2}$ are the photon energies and $r_{12}$ is the transverse distance
between their impact points on the calorimeter. We do the same for the remaining photon pair, which is labeled as  $B$. 
For a given pairing $(A,B)$, the calorimeter contribution to the $\chi^2_{\rm tot}$ is
defined as
\begin{align}
\chi_{\rm cal}^2(z_{\rm v})
= \frac{(z_A - z_{\rm v})^2}{\sigma_A^2}
+ \frac{(z_B - z_{\rm v})^2}{\sigma_B^2}.
\end{align}

The uncertainty $\sigma_A$ associated with $z_A$ is evaluated by standard error propagation using the photon energy and position resolutions, that is, 
\begin{align}
    \sigma_A^2 &= \left(\frac{\partial z_A}{\partial E_1}\right)^2 \sigma_{E_1}^2
    +\left(\frac{\partial z_A}{\partial E_2}\right)^2 \sigma_{E_2}^2
    +\left(\frac{\partial z_A}{\partial r_{12}}\right)^2 \sigma_{r_{12}}^2
    \\
    &=
    \frac{r_{12}^2}{4m_{\pi}^2}\left(
    \frac{E_2}{E_1}\sigma_{E_1}^2 + \frac{E_1}{E_2}\sigma_{E_2}^2
    \right)
    +\frac{E_1E_2}{m_{\pi}^2}\sigma_{r_{12}}^2
\end{align}
where the energy resolution $\sigma_{E}$ is in eq.~\eqref{eq:sigmaE}, and the position resolution is $\sigma_{r_{12}} =\sqrt{2}\sigma^{\mathrm{CsI}}$.
In a similar way,  $\sigma_B$ can be calculated. 

For each event, the four photons can be grouped into two pairs in three distinct ways.
Therefore, we have three different  $\chi^2_{\rm tot}$ functions and minimize each by fitting $z_{\rm v}$, and then the vertex position is obtained by the pair yielding the smallest $\chi^2_{\rm tot}$.  
Also, the photon pairing yielding the smallest $\chi^2_{\rm tot}$ is selected as the correct assignment, rather than the one minimizing $\chi_{\rm cal}^2$ alone.

\bibliography{ref}
\bibliographystyle{JHEP}
\end{document}